\documentclass[useAMS,usenatbib,]{mn2e}
\usepackage{times}
\usepackage{natbib}
\usepackage{amsmath}
\usepackage[dvips]{graphicx}
\usepackage{subfigure}
\title[]
{Measuring cosmic magnetic fields by rotation measure-galaxy cross-correlations in cosmological simulations}
\author[F. Stasyszyn, S. E. Nuza, K. Dolag, R. Beck, J. Donnert]
{F. Stasyszyn$^{1}$\thanks{E-mail: fstasys@mpa-garching.mpg.de}, S. E. Nuza$^{1}$, 
K. Dolag$^{1}$, R. Beck$^{2}$ and J. Donnert$^{1}$\\
\\
$^{1}$ Max-Planck-Institut f\"ur Astrophysik, Karl-Schwarschild Str. 1, D85748, Garching, Germany\\
$^{2}$ Max-Planck-Institut f\"ur Radioastronomie, Auf dem H\"ugel 69, 53121, Bonn, Germany}
\begin{document}

\date{Accepted. Received; in original form}


\maketitle

\label{firstpage}

\begin{abstract}
Using cosmological MHD simulations of the magnetic field in galaxy
clusters and filaments we evaluate the possibility to infer the magnetic 
field strength in filaments by measuring cross-correlation functions
between Faraday Rotation Measures (RM) and the galaxy density field. 
We also test the reliability of recent estimates considering the problem of 
data quality and Galactic foreground (GF) removal in current datasets. 
Besides the two self-consistent simulations of cosmological magnetic fields 
based on primordial seed fields and galactic outflows analyzed here, we also 
explore a larger range of models scaling up the resulting magnetic 
fields of one of the simulations.
We find that, if an unnormalized estimator for the cross-correlation 
functions and a GF removal procedure is used, the detectability of the 
cosmological signal is only possible for future instruments 
(e.g. SKA and ASKAP). However, mapping of the observed RM 
signal to the underlying magnetization of the Universe (both in space and time) 
is an extremely challenging task which is limited by the ambiguities of our 
model parameters, as well as to the weak response of the RM signal in low density 
environments. 
Therefore, we conclude that current data cannot constrain the amplitude and 
distribution of magnetic fields within the large scale structure and a detailed 
theoretical understanding of the build up and distribution of magnetic fields 
within the Universe will be needed for the interpretation of future observations.
\end{abstract}



\section{Introduction} 
\label{sec:intro}

Magnetic fields in the Universe are found in almost all studied environments. 
In particular, their presence in the inter-galactic medium \citep[IGM; see][ for a recent 
review]{2009ASTRA...5...43B} and in the intra-cluster medium (ICM) is confirmed by diffuse 
radio emission as well as by observations of Faraday Rotation Measures (RM) towards polarized 
radio sources within or behind the magnetized medium \citep[e.g.][]{2006AN....327..539G}. 
On the largest scales, like those of filaments, magnetic fields are notoriously difficult to measure and 
available data is still incomplete. This is especially difficult because these measurements require either a high thermal 
density (for RMs) or the presence of relativistic particles (for the synchrotron emission). 
Therefore, measurements of the magnetic field strength have been successfull for high density regions of collapsed objects 
(e.g. galaxies and galaxy clusters), and thus, fields significantly below the $\mu$G level can hardly 
be detected. 

Recently an interesting attempt to constrain the value of large scale cosmic magnetic fields was done 
by \citet{2009arXiv0906.1631L}. These authors detected a positive cross-correlation signal between 
the galaxy distribution in the SDSS Sixth Data Release \citep{2008ApJS..175..297A} and 
the RM values extracted from the \citet{2009ApJ...702.1230T} catalog. Using the amplitude 
of this signal, together with a simplified model for the magnetic fields configuration in the 
Universe (estimated from its mean electron density), and computing the RM typical 
values expected from this coherent field in a given length scale, they were able to derive 
limits for the corresponding cosmic magnetic fields. 

In this work, we want to investigate: ({\it i}) to what extent a self-consistent treatment of the 
cosmological RM signal based on magneto-hydrodynamical (MHD) simulations of structure formation changes the expected 
shape and amplitude of such a correlation signal, and ({\it ii}) how such an approach is affected by the presence of the 
Galactic foreground (GF) and noise in the final RM signal. Both points are of extreme importance, if 
robust field properties are to be derived from any observed signal. Furthermore, the appearance of 
magnetic field reversals (as observed in galaxy clusters at various length scales) will alter the cosmological
signal magnitude and shape, whereas the residuals of any foreground and measurement errors will
bias the relation between the amplitude of the correlation function and the underlying cosmological field.
In order to self-consistently treat the cosmic magnetic fields, we make use of several cosmological MHD simulations 
which compute the resulting magnetization of the cosmological structures (e.g. amplitude and structure) 
following different models for the origin and seeding process of such magnetic fields. We also 
construct magnetic field models with much higher magnetization amplitude in the low density regions
to test how the resulting signatures of more extreme models affect our results. Here we scale up the 
predicted amplitude of the magnetic field in filaments by several orders of magnitude to test if
such strong magnetic fields in low density regions significantly effect the expected correlation signal.  
By introducing GF and adding noise to the signal on top of the 
underlying cosmological signal, we can study how the shape and amplitude of the cross-correlation 
function would be modified when considering  actual observations. 
To avoid further complications we ignore the cosmological evolution
of magnetic fields, which, in principle, would be consistently treated within our cosmological
MHD simulations. Hence, we neglect the evolution of the cosmic magnetic field seen in the simulation 
as a result of the structure formation process, and assume the present day magnetization of the simulated 
universe to be present up to the redshift of the sources.

The paper is organized as follows. In Section \ref{sec:simul} we describe the cosmological 
MHD simulations used and how we compute the synthetic RM catalogs. 
In Section \ref{sec:cross} we discuss the cross-correlation estimators used, the estimation of 
the intrinsic uncertainties due to the limited number of lines of sight probing 
the magnetization of the cosmological structures, the different signals expected for 
the various magnetization of the universe, as well as the uncertainties induced by the 
redshift distribution of the sources. 
In Section \ref{sec:obs} we show how the shape and amplitude of the signal is affected by the recipe 
normally used to remove the foreground signal, due to observational noise and to the Galaxy itself. 
In Section \ref{sec:res} we summarize the combination of all the effects, and present the resulting 
observable signal of the different magnetic field models. Finally, our conclusions are given in 
Section \ref{sec:conc}.


\section{The Simulations}\label{sec:simul}

\subsection{The cosmological MHD simulations}
\begin{figure}
\begin{center}
\includegraphics[width=0.48\textwidth]{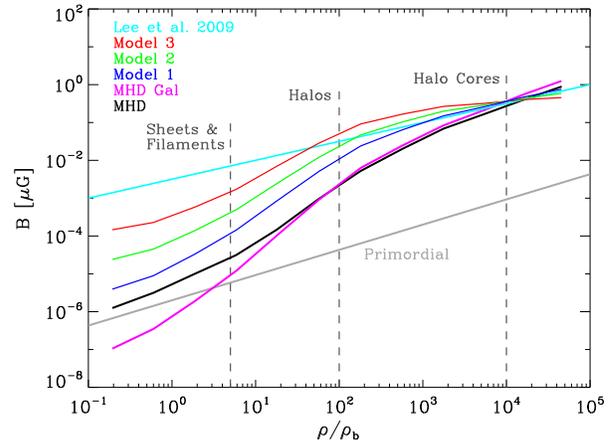}
\caption{Mean cosmic magnetic field as a function of density (in units of the mean cosmic baryon density) obtained from two, fully 
  self-consistent, cosmological MHD simulations for different magnetic field 
  origins ({\it MHD} and {\it MHD Gal}), as well as three models, where we 
  artificially scaled-up the magnetic field intensity at low densities to obtain scenarios 
  with extreme values in filaments ({\it Model 1}, {\it Model 2} and {\it Model 3}).
  For more details on these models see the text. Additionally, the primordial seed fields of 
  the {\it MHD} simulation and that obtained by \citet{2009arXiv0906.1631L} are shown.\label{fig:bmodels}}
\end{center}
\end{figure}

\begin{figure*}
\begin{center}
\includegraphics[width=0.8\textwidth]{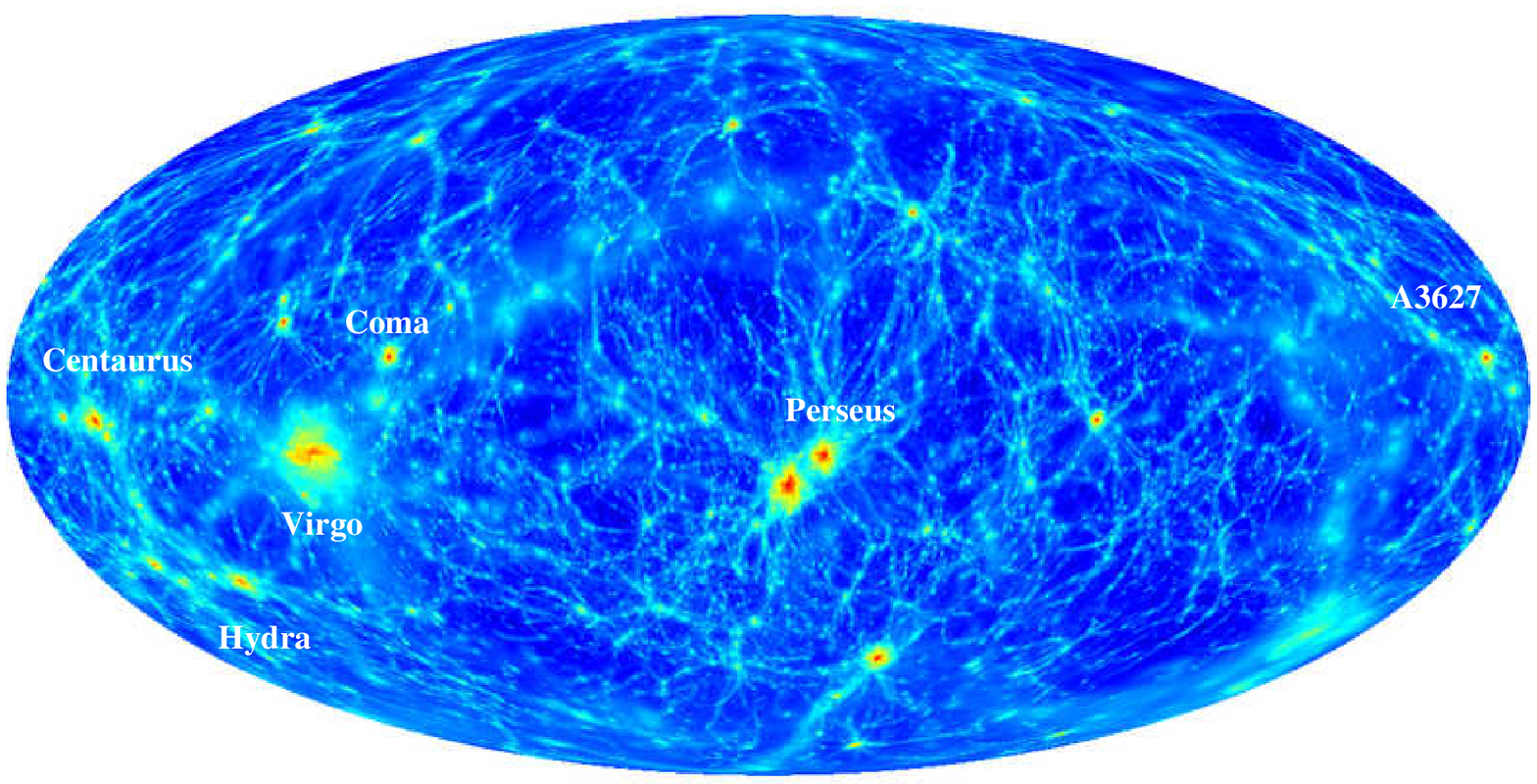}\\
\includegraphics[width=0.8\textwidth]{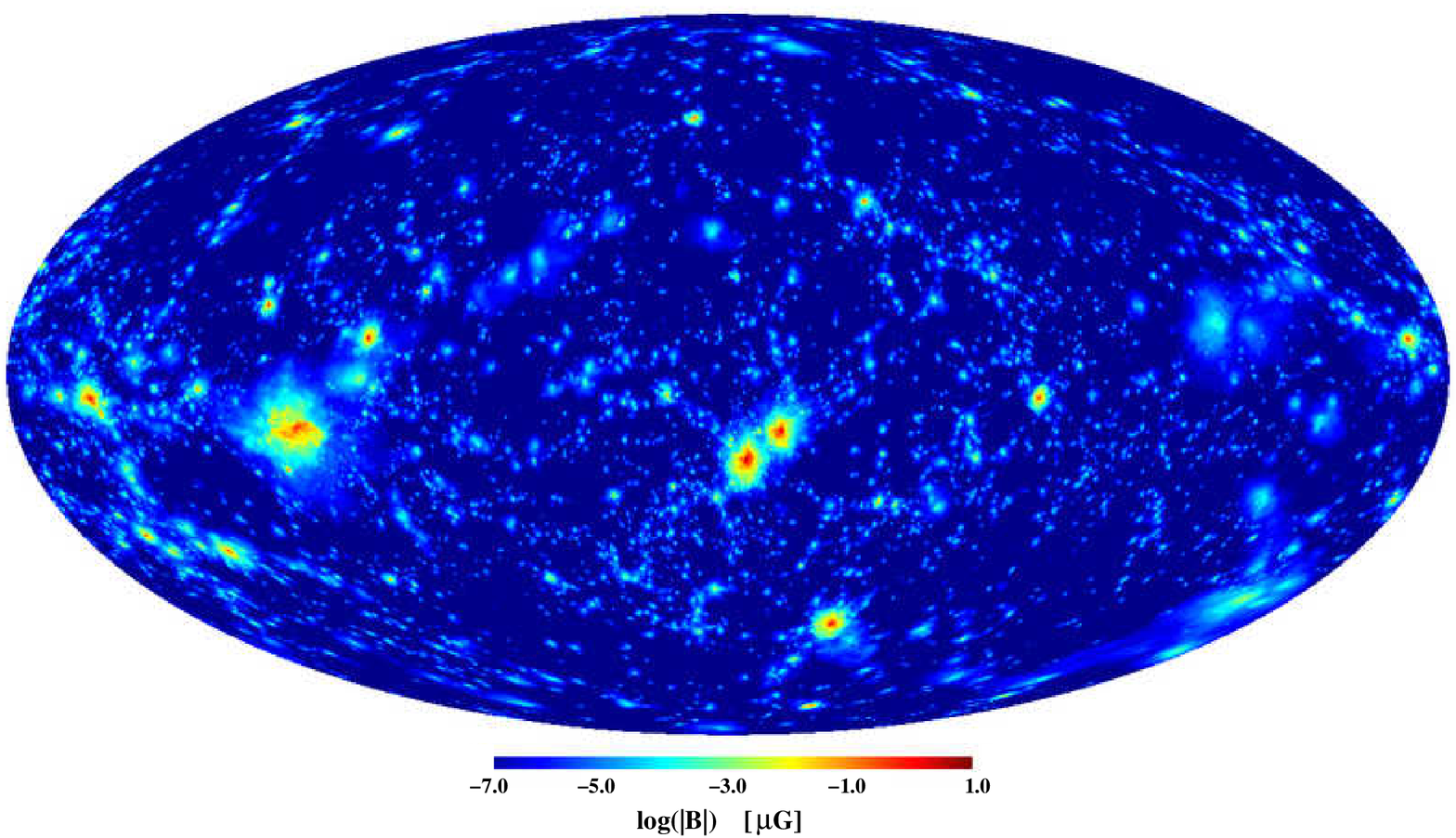}\\
\includegraphics[width=0.8\textwidth]{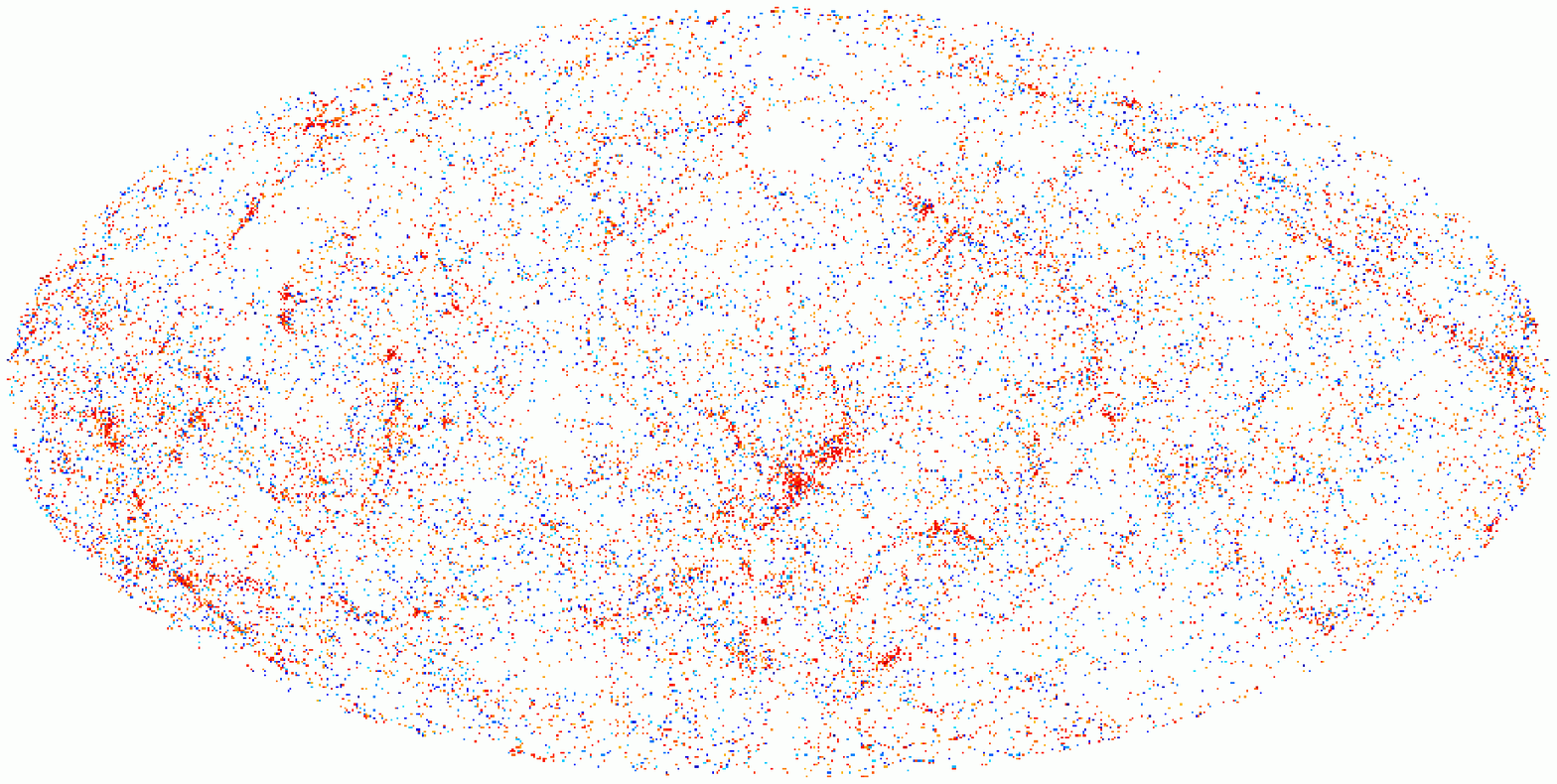}
\end{center}
\caption{Full sky maps of the local universe in supergalactic coordinates for the projected magnetic field in the {\it MHD} run (upper panel) 
and in the {\it MHD Gal} run (middle panel). The galaxy distribution expected from the corresponding hydrodynamical 
run is shown in the bottom panel. Galaxies are colour-coded from blue to red using their $B-V$ 
colours ($0.3<B-V<1$; see Nuza et al. 2010).\label{fig:simul}}
\end{figure*}

We used results from one of the constrained, cosmological MHD
simulations presented in \citet{2005JCAP...01..009D} and
\citet{2009MNRAS.392.1008D}.  In both simulations, the initial
conditions for a constrained realization of the local Universe were
the same as used in  \citet{2002MNRAS.333..739M}.  The 
initial conditions were obtained based on the the {\it IRAS} 1.2-Jy galaxy
survey \citep[see][ for more details]{2005JCAP...01..009D}. Its density
field was smoothed on a scale of $7\, \mathrm{Mpc}$, evolved back in
time to $z=50$ using the Zeldovich approximation, and used as an
Gaussian constraint \citep{1991ApJ...380L...5H} for an otherwise random
realization of a $\Lambda$CDM cosmology ($\Omega_M=0.3$,
$\Lambda=0.7$, $h=0.7$).  The {\it IRAS} observations constrain a volume of
$\approx 115 \, \mathrm{Mpc}$ centered on the Milky Way.  In the
evolved density field, many locally observed galaxy clusters  can be
identified by position and mass.  The original initial conditions were
extended to include gas by splitting dark matter particles into gas
and dark matter, obtaining particles of masses $6.9 \times 10^8\; {\rm
  M}_\odot$ and $4.4 \times 10^9\; {\rm M}_\odot$  respectively. The
gravitational softening length was set to $10\,\mathrm{kpc}$. 

\begin{table*} 
\begin{center}
\caption{Mean and standard deviations of RM absolute values for the
different catalogs at $z=0.03$ and estimates for $z=0.52$ and
$z=1.03$. The catalogs were constructed in each case using four
different realizations (see text). {\it MHD} is our fiducial structure
formation model. {\it Models 1, 2} and {\it 3} are its scaled-up 
versions. {\it MHD Gal} includes a semi-analytic model for Galactic
winds to seed magnetic fields at $z=4.1$. }
\begin{tabular}{lcccccc}
\hline 
\hline 
{\rm Model} & $\overline{|\rm{RM}|}_{z=0.03}$ & $\Delta\overline{|\rm{RM}|}_{z=0.03}$ 
            & $\overline{|\rm{RM}|}_{z=0.52}$ & $\Delta\overline{|\rm{RM}|}_{z=0.52}$ 
            & $\overline{|\rm{RM}|}_{z=1.03}$ & $\Delta\overline{|\rm{RM}|}_{z=1.03}$\\ 
            & [rad m$^{-2}$] & [rad m$^{-2}$]
            & [rad m$^{-2}$] & [rad m$^{-2}$] 
            & [rad m$^{-2}$] & [rad m$^{-2}$]\\ 
\hline 
{\it MHD Gal}    & 0.025 & 0.010 & 0.12 & 0.05 & 0.21 & 0.09 \\ 
{\it MHD}        & 0.018 & 0.010 & 0.09 & 0.05 & 0.15 & 0.09 \\ 
{\it Model 1}    & 0.018 & 0.008 & 0.09 & 0.04 & 0.15 & 0.07 \\ 
{\it Model 2}    & 0.025 & 0.010 & 0.12 & 0.05 & 0.21 & 0.09 \\ 
{\it Model 3}    & 0.040 & 0.013 & 0.20 & 0.06 & 0.34 & 0.11 \\ 
\hline 
\hline
\label{table1}
\end{tabular}
\end{center}
\end{table*}

The magnetic field was followed by our MHD simulations through the
turbulent amplification driven by the structure formation
process. For the magnetic seed fields, the first simulation (labeled 
{\it MHD}) followed a cosmological seed field (see Fig. \ref{fig:bmodels}), while in the 
second (labeled {\it MHD Gal}) we used a semi-analytic model for galactic winds. 
In particular, we considered the result of the {\it 0.1 Dipole} simulation from
\citet{2009MNRAS.392.1008D}. In both simulations, the resulting
magnetic field at $z=0$ reproduce the observed Rotation
Measure in galaxy clusters very well. A visual impression for the magnetic field
within the two different simulations and their corresponding galaxy distribution is 
shown in Fig. \ref{fig:simul}. 

\subsection{Artificial MHD models}

As is clearly visible in Fig. \ref{fig:simul}, such cosmological simulations 
usually predict relatively low magnetic fields in low density regions. To
explore more extreme models, we scaled up the magnetic field of the
{\it MHD} simulation by a factor
\begin{equation}
   B_{\rm 1,2,3} = B_{\rm MHD} \times
   \left(\frac{\rho}{\rho_{\rm scale}}\right)^\alpha,
\end{equation}  
with $\alpha$ being 1/3 ({\it Model 1}), 1/2 ({\it Model 2}) and 2/3
({\it Model 3}). Here ${\rho_{\rm scale}}$ denotes density scale for fixing  
the magnetic field, which we choose to be $10^4$ times the mean cosmic baryon density. 
The resulting behaviour of the mean magnetic field as a function of baryon 
density for the original runs, as well as for the scaled-up models, are
shown in Fig. \ref{fig:bmodels}. Note that the lines shown reflect 
the mean value of the magnetic field at the corresponding overdensity, while the 
dispersion of its amplitude can span several orders of 
magnitude in each density bin (see Dolag et al. 2005). We want to stress that
such scaled-up magnetic fields are artificial models, as the primordial field
needed to generate them would be well above current cosmological constraints 
(e.g. from CMB). Such strong seed fields would lead to an overprediction of 
the magnetic field amplitude in galaxy clusters by the simulations and it is 
quite unclear which physical process could be responsible to avoid this. 
We also remark that the scaled-up models lead to slightly lower central values for 
the magnetic field inside of galaxy clusters. This is qualitative in agreement with what 
is needed to fit the observed RM signal within the Coma galaxy cluster \citep[see][]{2010arXiv1002.0594B}.

\subsection{Synthetic RM catalogs}

For each of the 5 models, we construct full sky RM catalogs, sampling the whole sky
using 3072 different lines of sight (i.e. $\sim3^{\circ}$ resolution) making use
of the {\small HEALPix} \citep{2005ApJ...622..759G} tessellation of the sphere. Therefore, our RM
catalog contains roughly half as many number of lines of sight to probe the RM signal of 
the large scale structure than the catalog used as \citet{2009arXiv0906.1631L}.

Although we are only reproducing much shorter lines of sight than expected in the real Universe 
(due to the limited volume of the underlying simulation) we believe that the region probed 
reflects a fair representation of the present large scale distribution of 
galaxies \citep{next_nuza}, and therefore, we do not expect the amplitude of any normalized 
correlation signal to be strongly affected by a lack of fluctuation power. 

\subsection{Magnetic depth of the universe}

On its way to the observer, the polarized radio emission of the observed sources will
pass several times through the cosmological filamentary structures. The final RM value 
does accumulates in a random walk. Examples of the magnetic field structure
along some lines of sight through the simulated local universe can be found in previous work
(e.g. Dolag et al. 2005, Fig. 12; Dolag et al. 2009, Fig. 10). 
The magnitude of the observed RMs, and thus, the mean of the RM absolute 
values, will strongly depend on the {\it magnetic depth} (given by the redshift range probed) 
accessible to the observed sample of radio sources. In addition, if the magnetic field changes during 
the formation of the universe, such changes have to be convolved with the redshift
distribution of the observed sources.
For simplicity, we assume that the magnetization of the universe at the time of interest
was the same as today and that all sources towards the RMs are measured at the same
redshift, e.g. all lines of sight used probe the same {\it magnetic depth} of the universe. 

Unfortunately, our cosmological MHD simulation is much smaller than is required to compare 
with observations directly and therefore we have to extrapolate our calculated RMs to the 
redshift of the real observed sources. With the size of our simulation box, we can probe only out to $z=0.03$. 
Because of this, we account for the increase of the RM values due to a random walk process towards 
higher redshift by assuming the same contribution of cosmic structures to estimate the cosmological RMs. This is 
done by replicating the original volume 15 and 22 times (corresponding to redshifts out to $z=0.52$ and $z=1.03$). 
As shown in Table \ref{table1}, the associated RM amplification factors, including the shift of the 
rest-frame frequencies due to the cosmological expansion for each replication of the box, are 4.97 and 8.54 respectively.
We will use such expected amplified RM signals in the following analysis, 
indicating this by adding the redshift used for the {\it magnetic depth} together 
with the model name. Note that to increase the RM signal by a factor of 100 one must 
probe cosmic structures up to $z\approx8$.

Even for the extreme scaled-up models and extrapolation 
out to $z=1.03$, the expected RM signal is still one order of magnitude smaller than 
the reported value by \citet{2009arXiv0906.1631L} in their simplified 
model (i.e. $|{\rm RM}| \approx 2$ rad m$^{-2}$). This emphasizes the fact that 
simulations that properly take the cosmological structures into account are needed 
to relate any possible correlation signal to global magnetic field 
values. Note also that such small signals are expected to be very sensitive to measurement errors which will 
scale with the even much larger foreground signal imposed by our galaxy. We explore these problems in 
the following sections. 


\section{Evaluating the cosmological Cross-Correlation Signal} \label{sec:cross}

\subsection{Estimators}

\begin{figure*}
\includegraphics[width=0.46\textwidth]{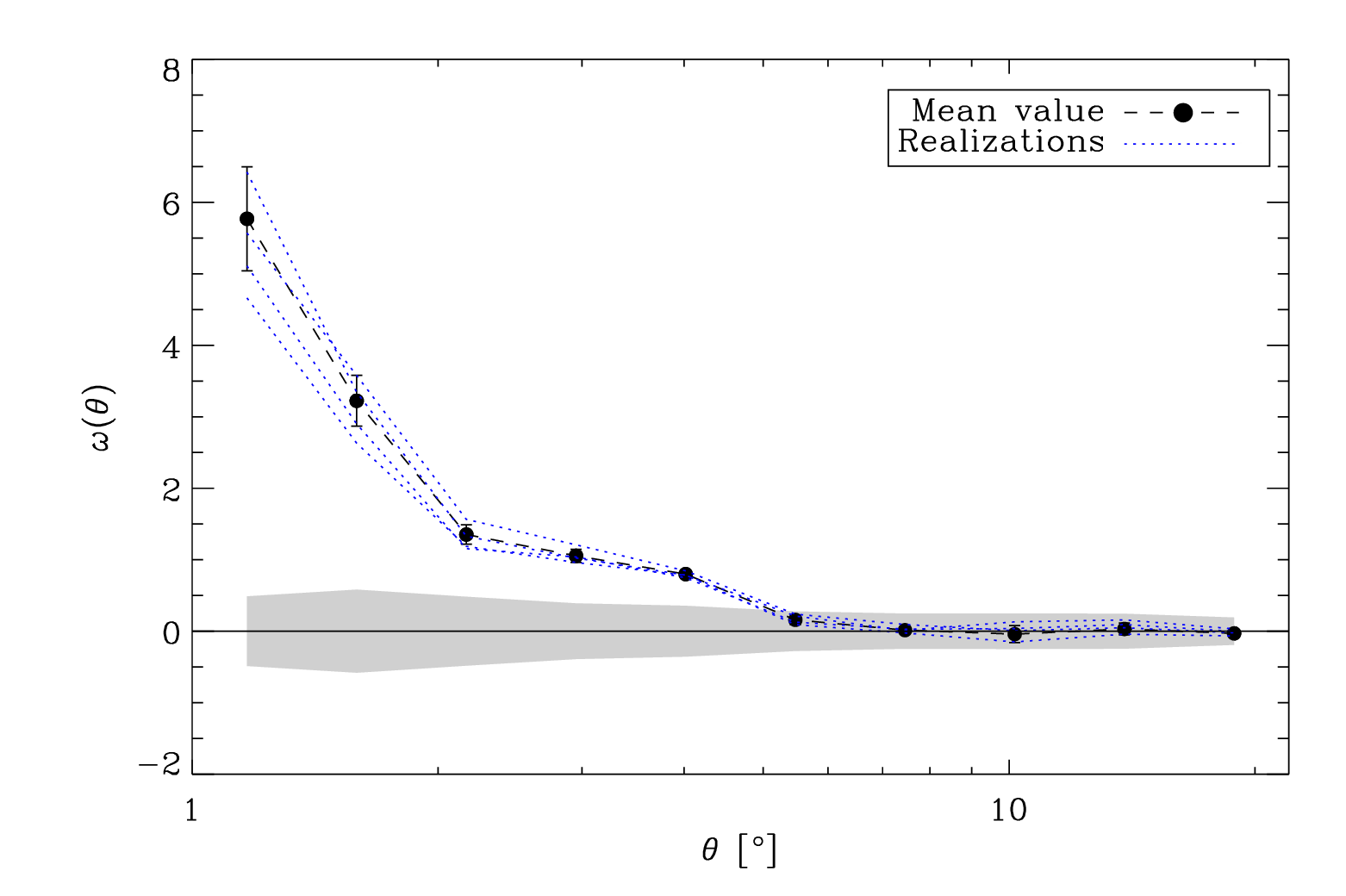}
\includegraphics[width=0.46\textwidth]{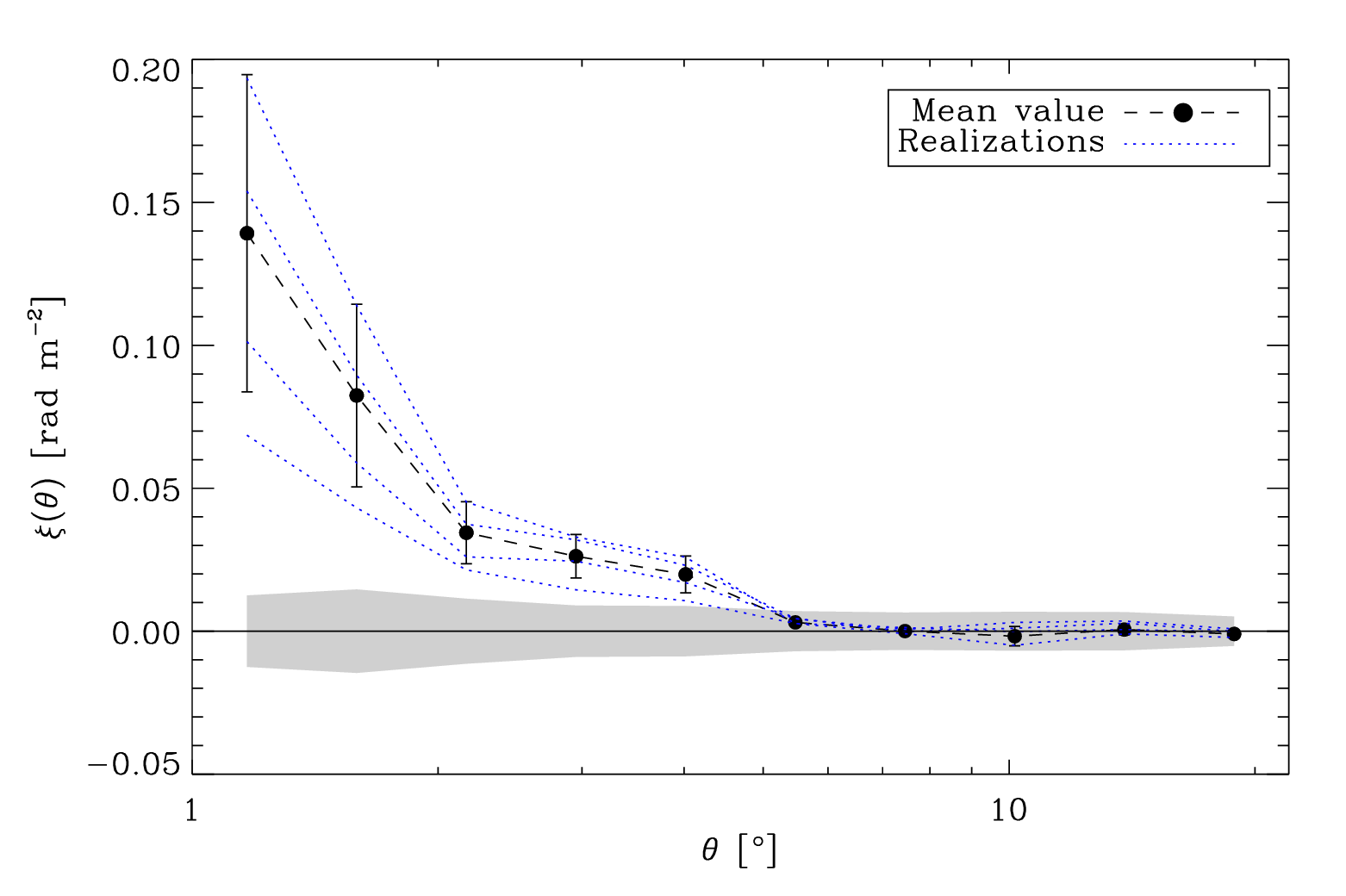}
\caption{Angular cross-correlation functions, for the full sky map based on the 
original {\it MHD Gal} simulation, using the two estimators presented in the 
text (see Eqs. \ref{omega_est} and \ref{xi_est}). In both panels, the results of 
different magnetic field realizations can be seen as blue dotted lines. The average result is 
shown as black filled circles. Error bars denote the 1-$\sigma$ dispersion 
due to the different realizations. The grey area indicates the {\it null} signal obtained by 
reshuffling the RMs.\label{fig:cross_models}}
\end{figure*}

We compute the cross-correlation signal between the RM computed along 3072 
lines of sight using a {\small HEALPix} tessellation of the sky
and the angular positions of simulated galaxies. For every
direction, we count the number of galaxies lying at an angle between
$\theta$ and $\theta+d\theta$, weighting the counts with the
corresponding absolute RM value. Formally, the cross-correlation
function between $|\rm{RM}|$ and the galaxy density $n$ is defined as
follows

\begin{equation}
\omega_{\rm RM}(\theta) \equiv \frac{\langle\Delta n(\theta)|\rm{RM}|\rangle}{\bar{n}\rm{\overline{|RM|}}}\rm{,}
\label{omega_est}
\end{equation}  

\noindent where $\Delta n$ measures the fluctuations around the mean
value of $n$, $\bar{n}$ is the mean density of the galaxy sample,
${\rm{\overline{|RM|}}}$ is the mean of the $|\rm{RM}|$ catalog, and
$\langle\ldots\rangle$ denotes ensemble average. If the 
distributions of $n$ and RM are Gaussian, this estimator is 
insensitive to the addition of an uncorrelated signal (like noise 
and/or foreground). While for the galaxy density $n$ a Gaussian distribution is still
a reasonable assumption, the RM absolute values are strongly non-Gaussian. 
Therefore, it cannot be easily predicted how this estimator 
will behave once the observational process is included. 
In fact, \cite{2009arXiv0906.1631L} used a different estimator, where
the normalization by $\overline{|\rm{RM}|}$ is not evaluated, i.e.

\begin{equation}
\xi_{\rm RM}(\theta) \equiv \frac{\langle\Delta n(\theta)|\rm{RM}|\rangle}{\bar{n}}\rm{.}
\label{xi_est}
\end{equation}  

\noindent Note that this estimator is likely quite sensitive 
to processes which change the value of $\overline{|\rm{RM}|}$, such as the {\it magnetic
depth} probed by the redshift distribution of the sources.

\subsection{Evaluating uncertainties}

We investigate two kind of errors for our simulated cross-correlation functions.
To estimate the significance of our obtained correlation signal, we randomly 
shuffled the RM data twenty times. This allow us to take into 
account the variance given by the used sampling and reveals the significance of 
the correlation itself. We will indicate this as the expected level for a {\it null} 
signal in the figures. A second source of errors is given by the magnetic field realizations 
available inside the simulated volume. Since we are still 
using a small number of RM points (i.e. 3072), this also introduces significant noise to the
correlation functions obtained. It is beyond the scope of the present work to produce several 
independent simulations based on different realizations of the initial magnetic seed field.
Therefore, we assesed this by calculating the RM signal of each particle using 
either the $x$, $y$ and $z$ component or the radially projected 
magnetic field component. Note that the first three components are only statistically equivalent 
to the radially projected component once a isotropic distribution of the magnetic field is assumed. 
Next we use these four realizations of the RM signal to estimate 
the uncertainty according to the underlying cosmic magnetic field realization in the simulations.
This uncertainty is then added as error bars to the individual models.

Fig. \ref{fig:cross_models} shows the correlation function signal obtained from 
the {\it MHD Gal} model for the two estimators ($\omega_{\rm RM}$ and $\xi_{\rm RM}$). 
It shows the individual signal for the four different realizations of the magnetic fields
and the resulting mean signal with their corresponding error bars. The {\it null} signal 
obtained from reshuffling the RMs twenty times is indicated by the grey area. 
Whereas the uncertainties coming from the magnetic field variance in the different 
realizations changes the amplitude of $\omega_{RM}(\theta)$ only by 10\%,
it is clearly visible that using the unnormalized $\xi_{RM}(\theta)$ estimator 
introduces much larger uncertainties in its amplitude (by roughly a factor $\sim2$). 
It also highlights the fact that using this estimator together with the mean of the 
$|{\rm RM}|$ signal to infer the underlying magnetic field will generate large 
uncertainties in the estimation. 
It is important to keep in mind that the ratio between both estimators (in every scale) 
for each of the individual realizations is given by the mean value of the corresponding 
absolute RMs. However, the normalization of the different realizations of each magnetic 
field model can vary significantly (up to a factor of $\sim3$). Whereas in the normalized
cross-correlation this is naturally absorbed, in the unnormalized case it enters in the error 
bars when building the assembly average over different realizations.

As a final remark, we expect that both errors will decrease 
similarly as Poissonian errors do when the number of lines of sight to probe the RM signal 
is increased.

\subsection{Magnetic depths}

\begin{figure*}
\includegraphics[width=0.46\textwidth]{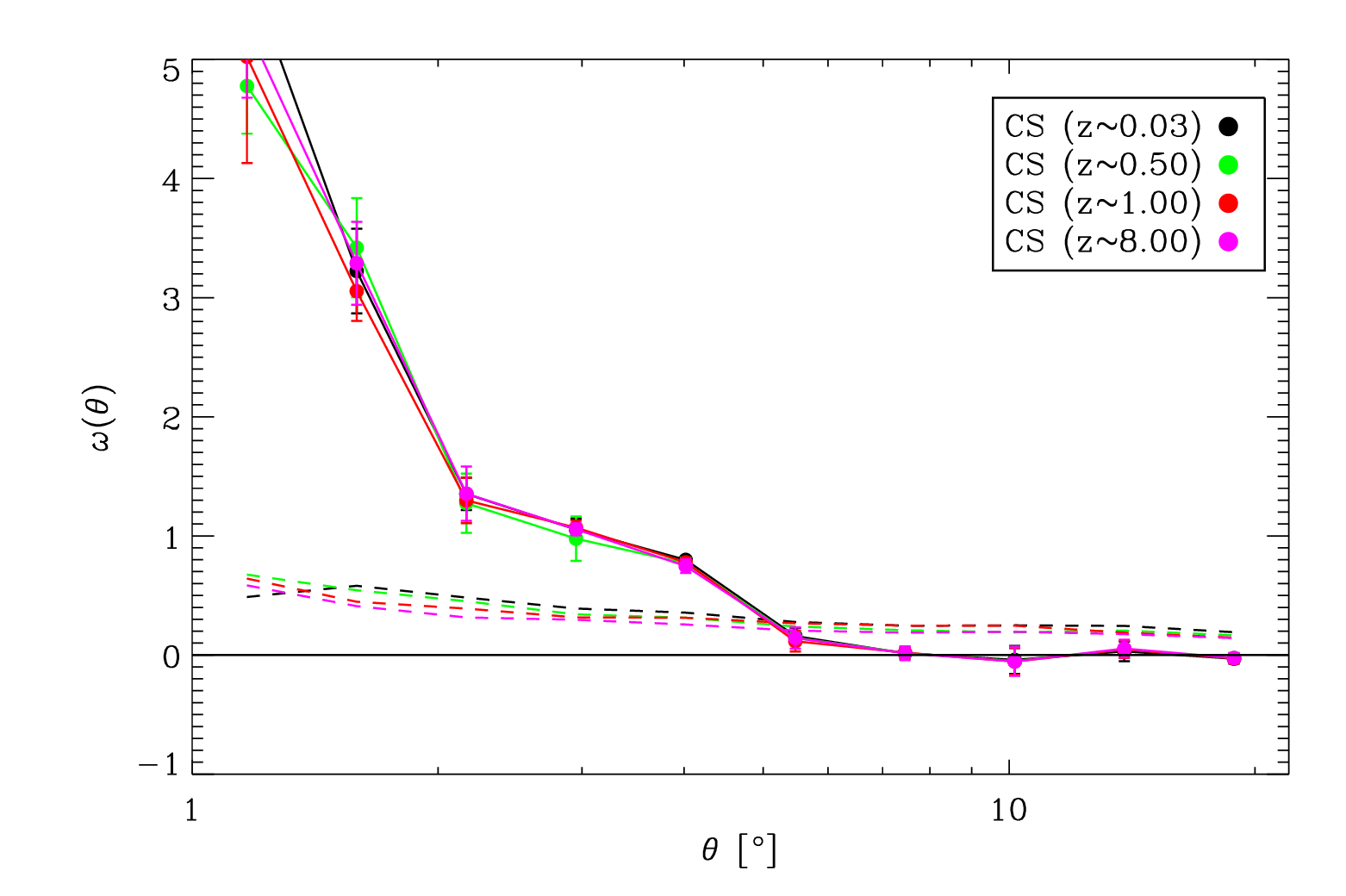}
\includegraphics[width=0.46\textwidth]{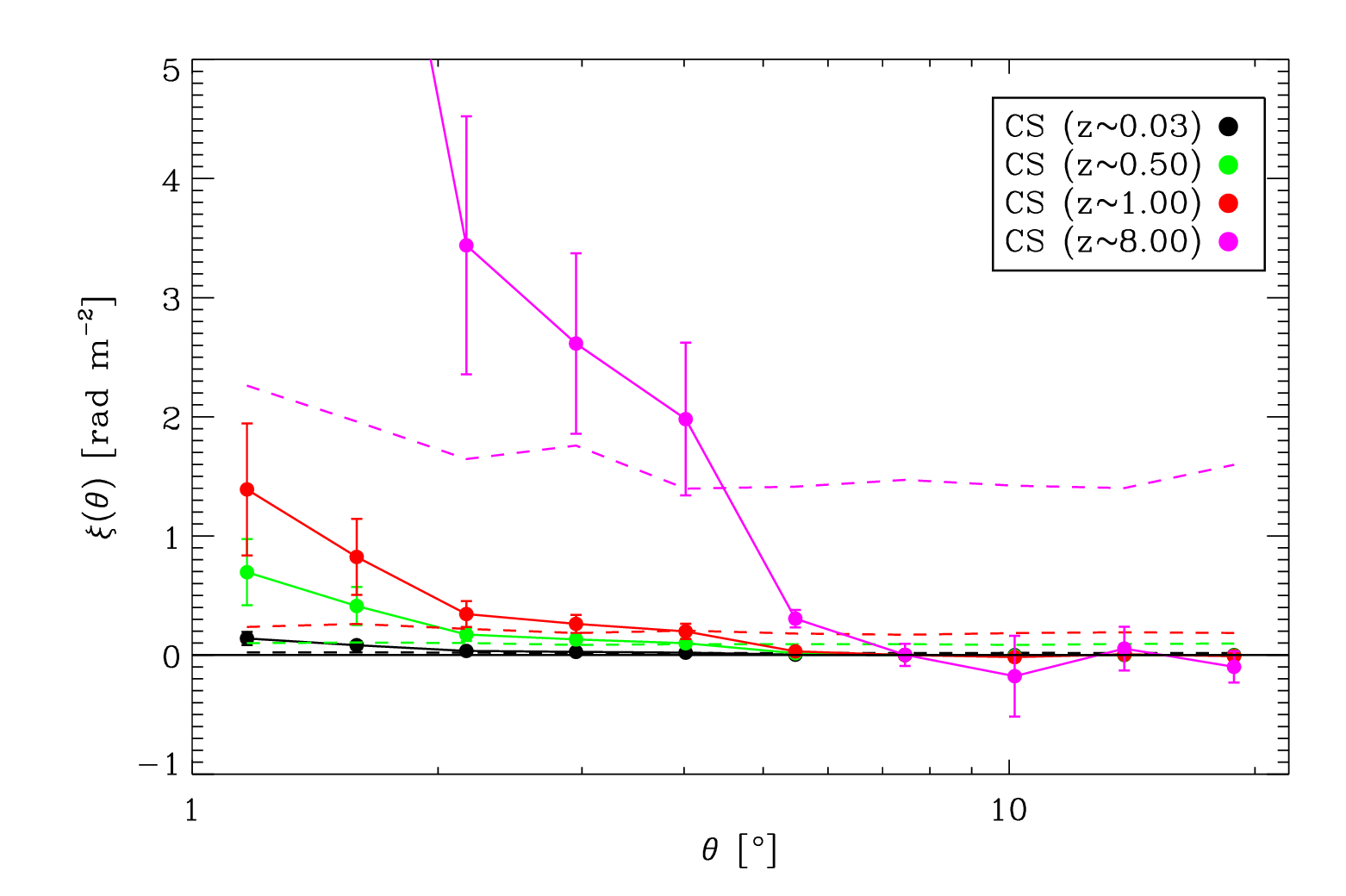}
\caption{Angular cross-correlation function, based on the original {\it MHD Gal} cosmological signal (CS),
evaluated for different {\it magnetic depths} probed by the RMs, using the two estimators 
presented in the text (left and right panels).\label{fig:cross_depth}}
\end{figure*}

As mentioned before, to correctly interpret the cross-correlation signal we must consider 
the {\it magnetic depth} probed by the RMs. 
Fig. \ref{fig:cross_depth} shows the expected signal, assuming a non-evolving magnetic 
field probed by the sources up to a certain given redshift. 
As expected, the normalized correlation function signal $\omega_{RM}(\theta)$ is 
insensitive to the particular probed volume and its amplitude reflects the 
underlying magnetic field distribution, independent on the redshift distribution
of the sources. On the other hand, the signal given by the unnormalized estimator $\xi_{RM}(\theta)$ 
increases as a function of the {\it magnetic depth} as expected. In this case, the amplitude changes
by more than a factor of two if the redshift distribution of the radio sources is changed from
$z=0.52$ to $z=1.03$. This means that one has to consider the redshift distribution 
of the radio sources towards the observed RMs in order to relate the 
amplitude of the signal to the underlying magnetization of the large scale 
structures. Therefore, it is difficult to interpret such an observed signal \citep[as done e.g. by][]{2009arXiv0906.1631L}. 

\subsection{Magnetic field models}

\begin{figure*}
\includegraphics[width=0.46\textwidth]{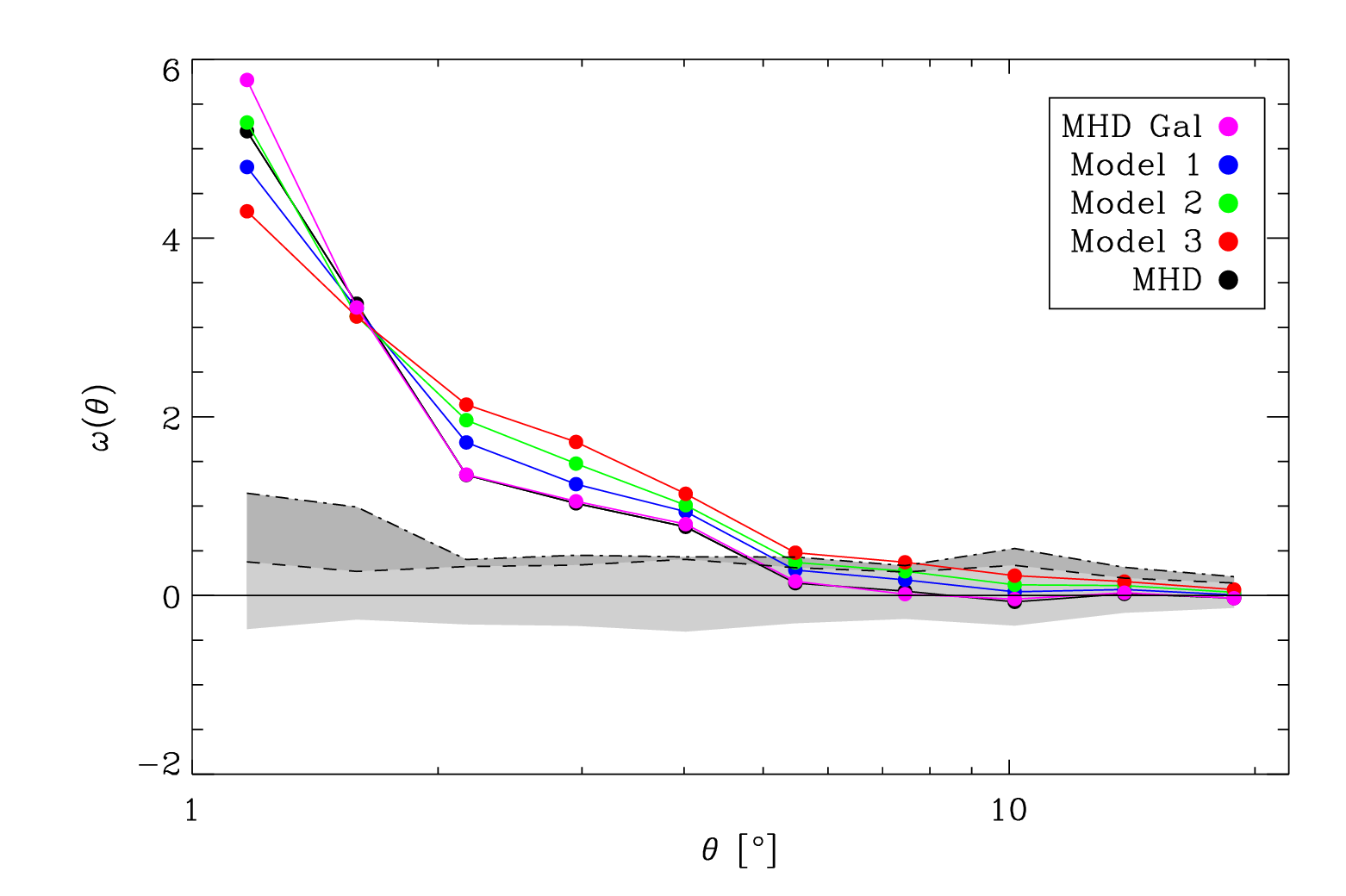}
\includegraphics[width=0.46\textwidth]{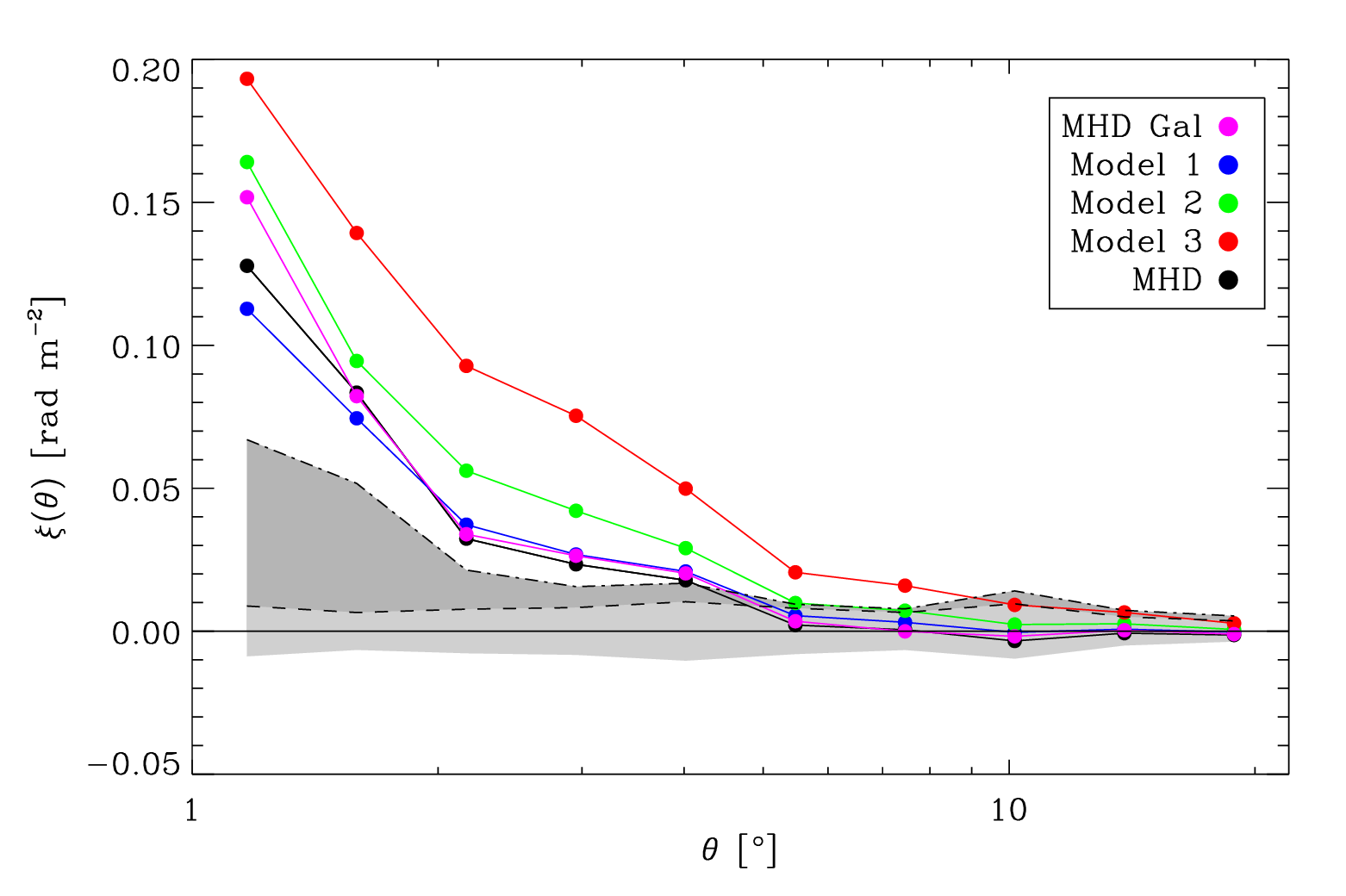}
\caption{Comparison between the angular cross-correlation functions,
  for the full sky maps in different models, using the two estimators
  presented in the text (left and right panels). Black solid line
  indicates the {\it MHD} model, while blue, light green and red
  indicate {\it Model 1, 2} and {\it 3} respectively. Pink solid
  line indicates the {\it MHD Gal} model. In both panels, the light grey
  shaded area indicates the randomly shuffled region, while the
  dark grey area indicates the magnitude of typical errors present 
  at a given scale due to the different RM realizations.\label{fig:cross_models2}}
\end{figure*}

Fig. \ref{fig:cross_models2} shows the cross-correlation function signal
obtained from the five different models investigated. The two shaded 
regions indicate the contribution of the two errors discussed 
before. The shape and the ordering of the 
correlation signal of $\omega_{RM}$ reflects the scaling of 
the underlying magnetic field models with density. In particular, the crossover
of the correlation function reflects the one seen in the underlying
magnetic field models very well (see Fig. \ref{fig:bmodels}). The 
correlation signal thus indeed carries information about the strength and distribution
of the cosmic magnetization. It is expected that using more 
lines of sight will reduce the statistical errors enough to make our
extreme models clearly distinguishable. This would be in principle 
possible with the available number of line of sights in 
current data \citep[e.g.][]{2009ApJ...702.1230T}.

The unnormalized correlation function $\xi_{RM}$ leads
to a larger relative change of the amplitude of the correlation signal
for the different magnetic field models, especially for the ones with 
very high magnetic fields in filaments. However, the resulting signal
comes with much larger errors (coming mainly from the different magnetic field 
realizations in the same models) and is therefore less significant. 
Also, the ordering of the magnetic field models with less extreme magnetic 
field values in filaments is not longer reflected in the correlation 
amplitude, particularly towards smaller impact parameters.

In summary, we conclude that the correlation signal for $\omega_{RM}$
inherits a clear signal from the cosmological magnetization, whereas the
correlation function $\xi_{RM}$ \citep[as used in][]{2009arXiv0906.1631L} 
is very difficult to interpret. Because of the missing normalization,
changes in the underlying RM distribution (caused by different realizations 
of the same magnetic field model or the {\it magnetic depth} probed by the 
radio sources) are not compensated for.


\section{Simulating the observational process} \label{sec:obs} 

To test for the effects caused by the observational process on the cross-correlation functions 
it is important to use an underlying scenario which reflects the expected amplitude of the RM signal. 
Therefore, we use (unless specifically stated) an underlying {\it magnetic depth} of the universe of $z=1$.

\subsection{Foreground removal procedure}
\label{for_rem}

\begin{figure}
\includegraphics[width=0.46\textwidth]{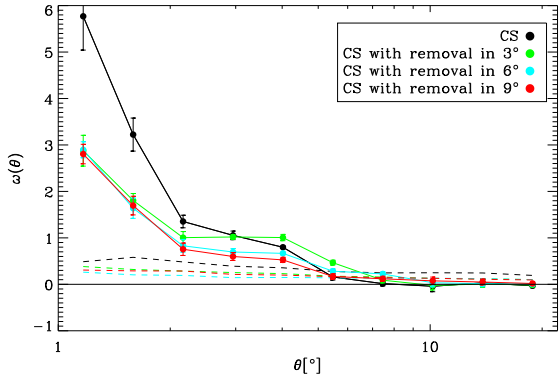}
\caption{Angular cross-correlation function, based on the {\it MHD Gal} cosmological signal (CS), 
  assuming a magnetic depth of $z=0.03$. Shown are the results obtained by 
  subtracting to each $|{\rm RM}|$ the average of its neighbors (excluding itself) within a radius 
  of 3$^{\circ}$, 6$^{\circ}$ and 9$^{\circ}$ for the $\omega_{RM}$ estimator.\label{fig:cross_agv2}}
\end{figure}

For RM observations, the removal of the foreground imposed by our galaxy (GF) is a major problem. 
Usually one assumes that the foreground varies on (much) larger scales than the ones of interest 
and removes the GF by subtracting a smoothed signal from the original data. 
Here we test how such a removal procedure affects the underlying cosmological signal 
traced using correlation functions following exactly the same procedure applied 
in \citet{2009arXiv0906.1631L}. At every point, we subtract the mean of the RM absolute values 
within a given radius (excluding the central value). Specifically, we tested three different angular 
sizes for the removal (i.e. $3^{\circ}$, $6^{\circ}$ and $9^{\circ}$). Fig. \ref{fig:cross_agv2} shows the result of such 
foreground subtraction technique on the normalized correlation function for the cosmological signal
using the normalized estimator $\omega_{RM}$. At small distances, this procedure leads to
a significant suppression of the correlation signal, even up to a factor of $\sim2$ for angular 
distances below $\sim2^{\circ}$, almost independently of the size of the removing radius. At larger 
distances, the amplitude of the correlation function is slightly increased (10 -- 20\%) starting from scales 
larger than the smoothing radius.

\subsection{Adding observational noise}

\begin{figure}
\includegraphics[width=0.46\textwidth]{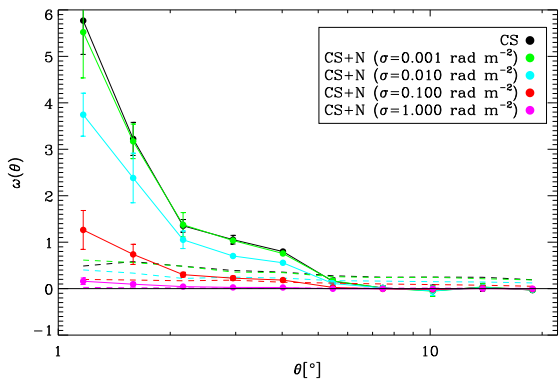}
\caption{Angular cross-correlation function for the {\it MHD Gal } cosmological signal (CS) using the $\omega_{RM}$ estimator. 
Same as Fig. \ref{fig:cross_models} (left panel), but including the effects of different random noise (N) scenarios at 
magnetic depth of $z=0.03$.\label{fig:cross_agv3}}
\end{figure}

Another problem for the observed RMs are the measurement errors by themselves. 
For example, since the data recently published by \citet{2009ApJ...702.1230T} is based on only 
two different frequency bands the resulting RMs will be affected by a significant uncertainty. 
We also note that these errors are not reduced by the smoothing involved when removing the  
foreground. The typical error of the observational RMs (as inferred from comparison with a data subset 
which was observed at more frequency bands) turns out to be around 10 to 20 rad m$^{-2}$ 
\citep[see][ Fig. 2]{2009ApJ...702.1230T}. In order to estimate the effect of the observational errors, we 
added random values to our simulated RM signal, which were drawn from a Gaussian distribution with a dispersion given 
by $\sigma_{\rm RM}$. We explored values of $0.001$, $0.01$, $0.1$, $1.0$ and $10$ rad m$^{-2}$ for $\sigma_{\rm RM}$.
Note that most of these values are much more optimistic than what is expected from current instruments. 
However, future instruments, like e.g. SKA and ASKAP will achieve an RM accuracy of a 
few rad m$^{-2}$ \citep{2004NewAR..48.1289B}.

Fig. \ref{fig:cross_agv3} shows the impact of such measurement errors onto the resulting
correlation function. Even $\sigma_{\rm RM}=0.01$ (e.g. a hundredth of the actual 
measurement error) leads to a sizable (ca. 50\%) reduction of the correlation signal. 
Furthermore, $\sigma_{\rm RM}=0.1$ (e.g. ten percent of the actual measurement error) reduces
the signal by a factor of $\sim5$ and $\sigma_{\rm RM}=1$ (e.g. nearly the present measurement errors) 
makes the correlation very close to the one of the corresponding {\it null} signal. 
From this it is clear that using the normalized estimator $\omega_{RM}(\theta)$ 
will be quite problematic. The presence of even small measurement errors (far smaller than what 
can be reached currently) will affect the shape and amplitude of the correlation function in a way 
that the information on the cosmic magnetization is basically lost.
 
\subsection{Adding Galactic foreground}
\label{hammu}

\begin{figure*}
\subfigure[CS+GF+N($\sigma=10.0$ rad m$^{-2}$)]
	{\includegraphics[width=0.49\textwidth]{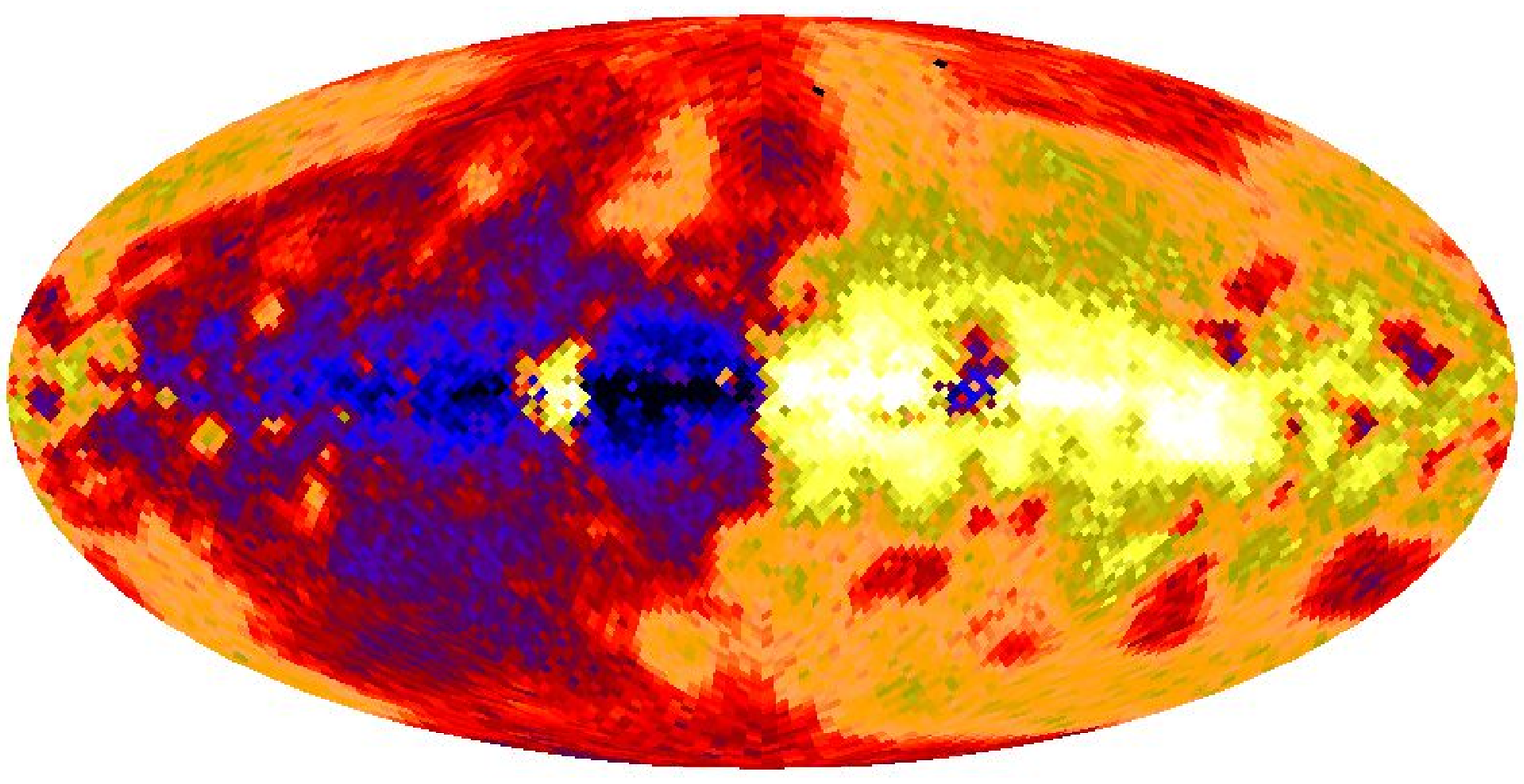}}
\subfigure[RM data]
	{\includegraphics[width=0.49\textwidth]{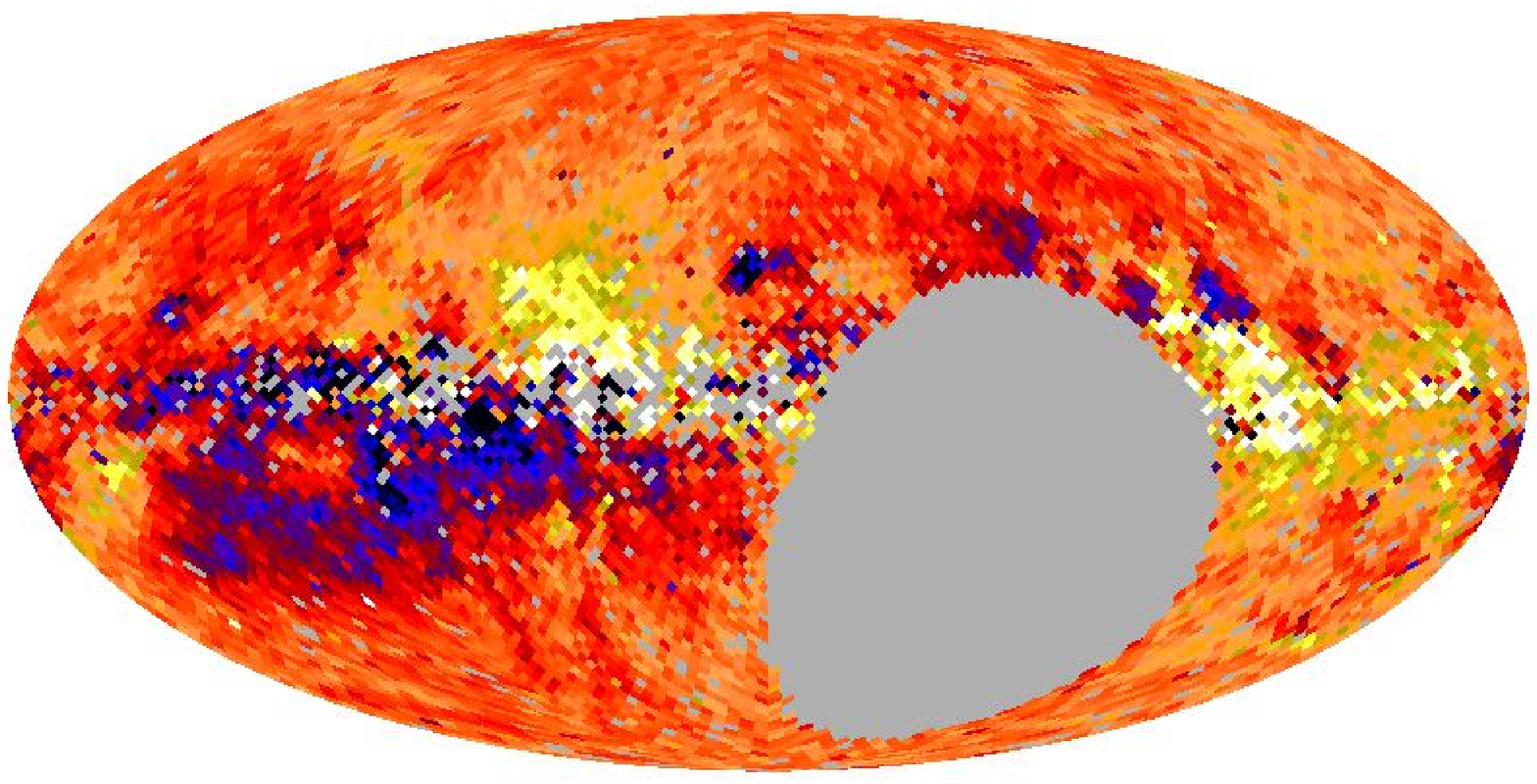}}\\
\subfigure[CS+GF+N($\sigma=10.0$ rad m$^{-2}$) smoothed in $8^\circ$]
	{\includegraphics[width=0.49\textwidth]{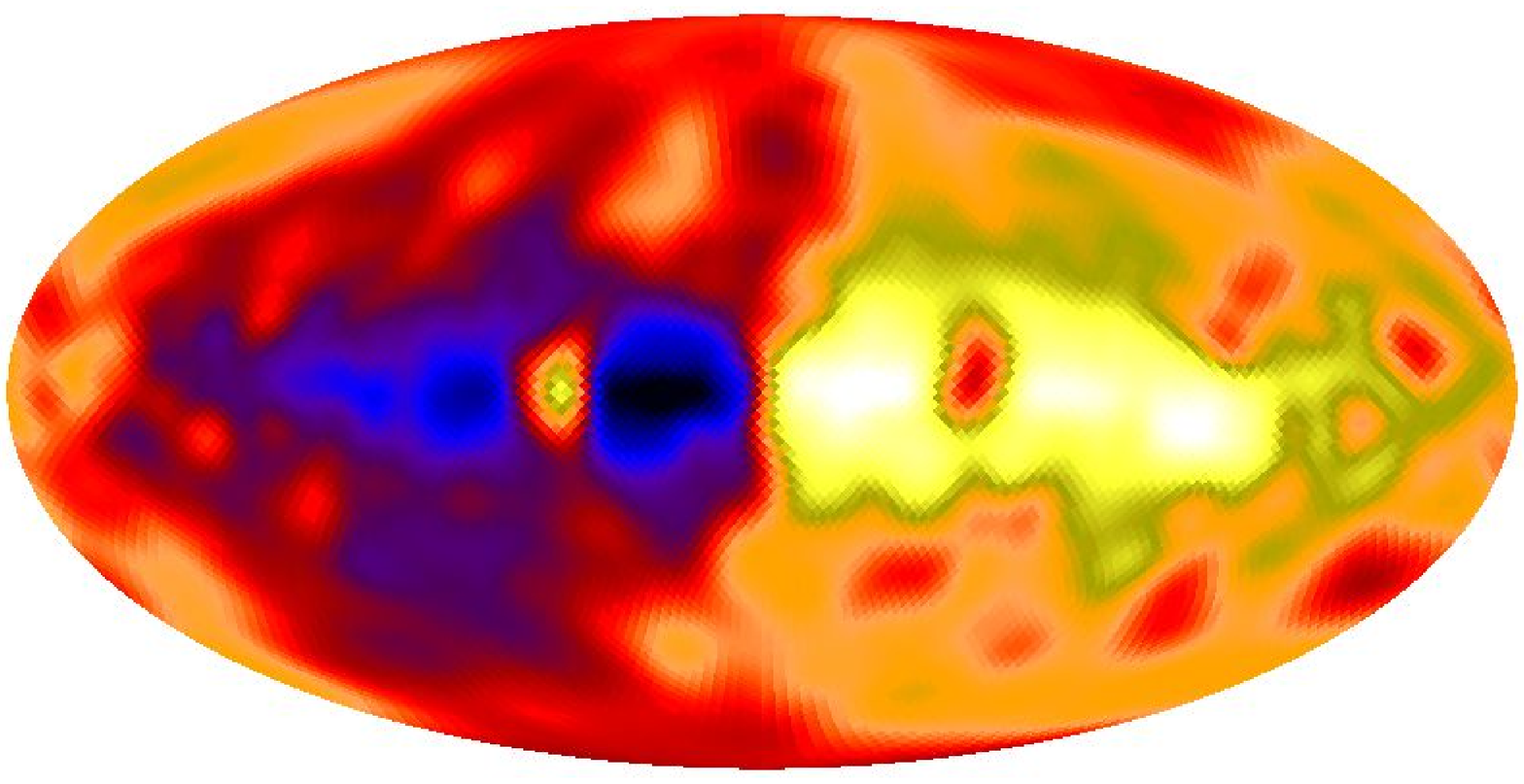}}
\subfigure[RM data smoothed in $8^\circ$]
	{\includegraphics[width=0.49\textwidth]{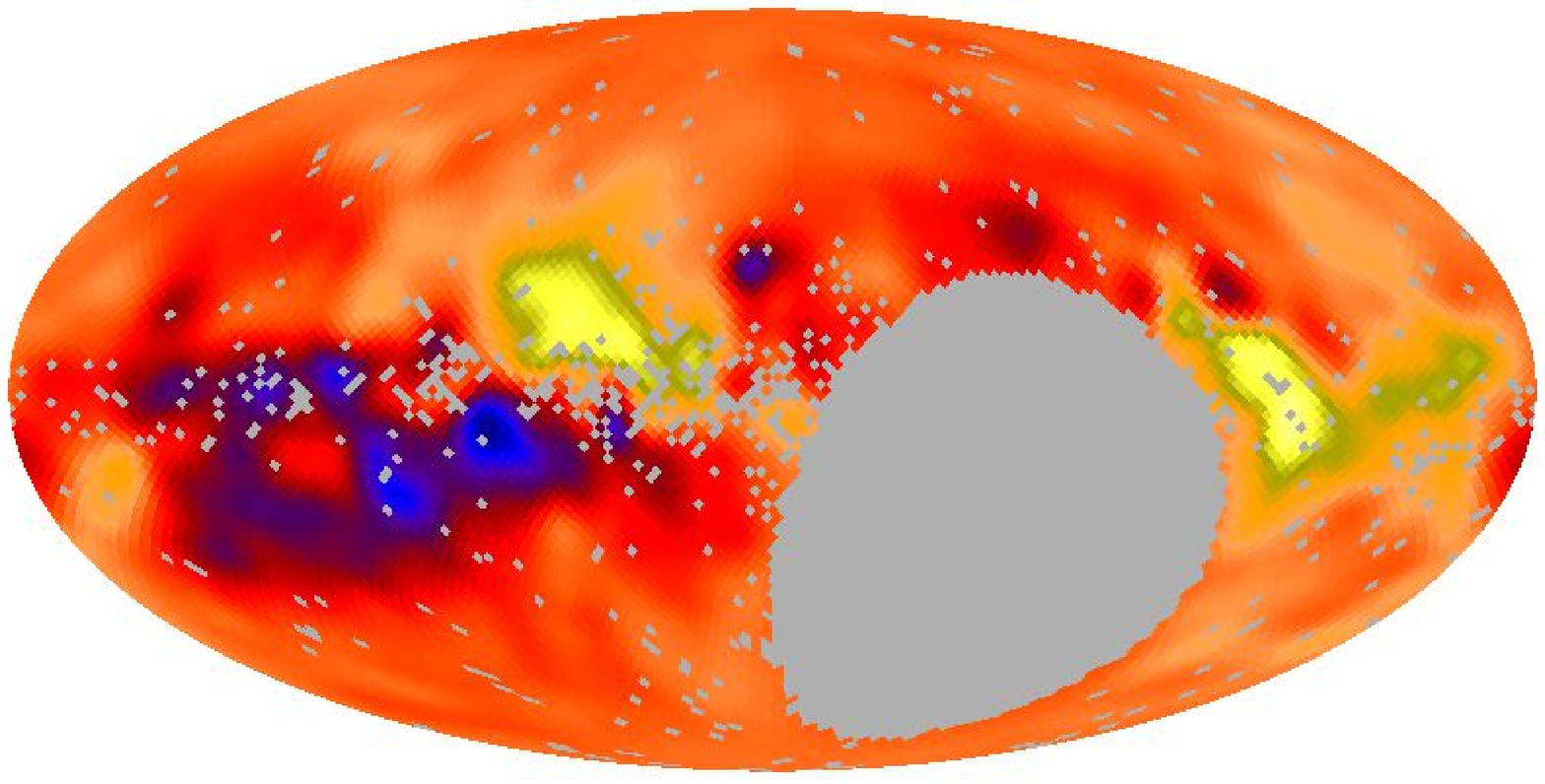}}\\
\subfigure[CS+GF+N($\sigma=10.0$ rad m$^{-2}$) with GF removal in $3^\circ$]
	{\includegraphics[width=0.49\textwidth]{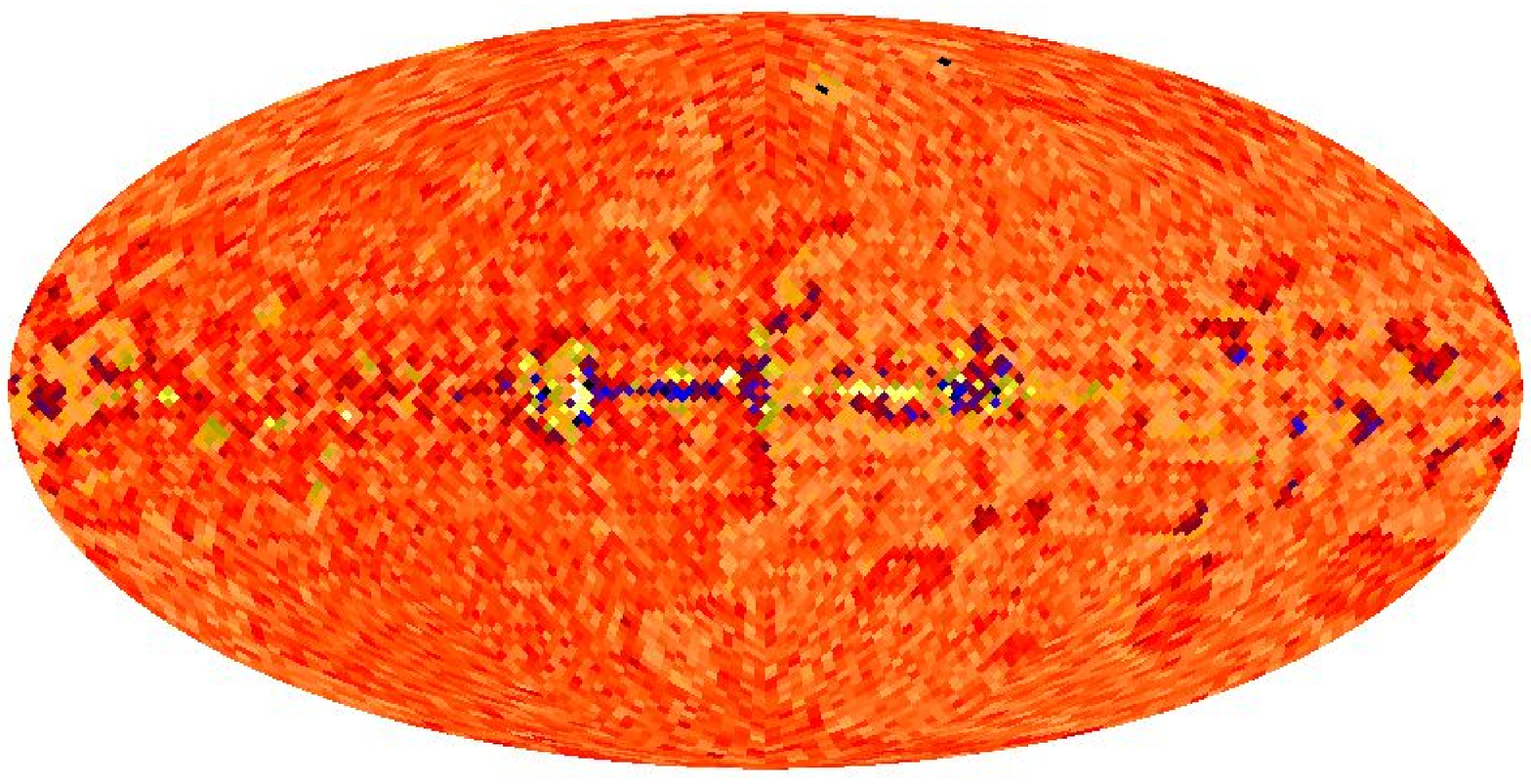}}
\subfigure[RM data with GF removal in $3^\circ$]
	{\includegraphics[width=0.49\textwidth]{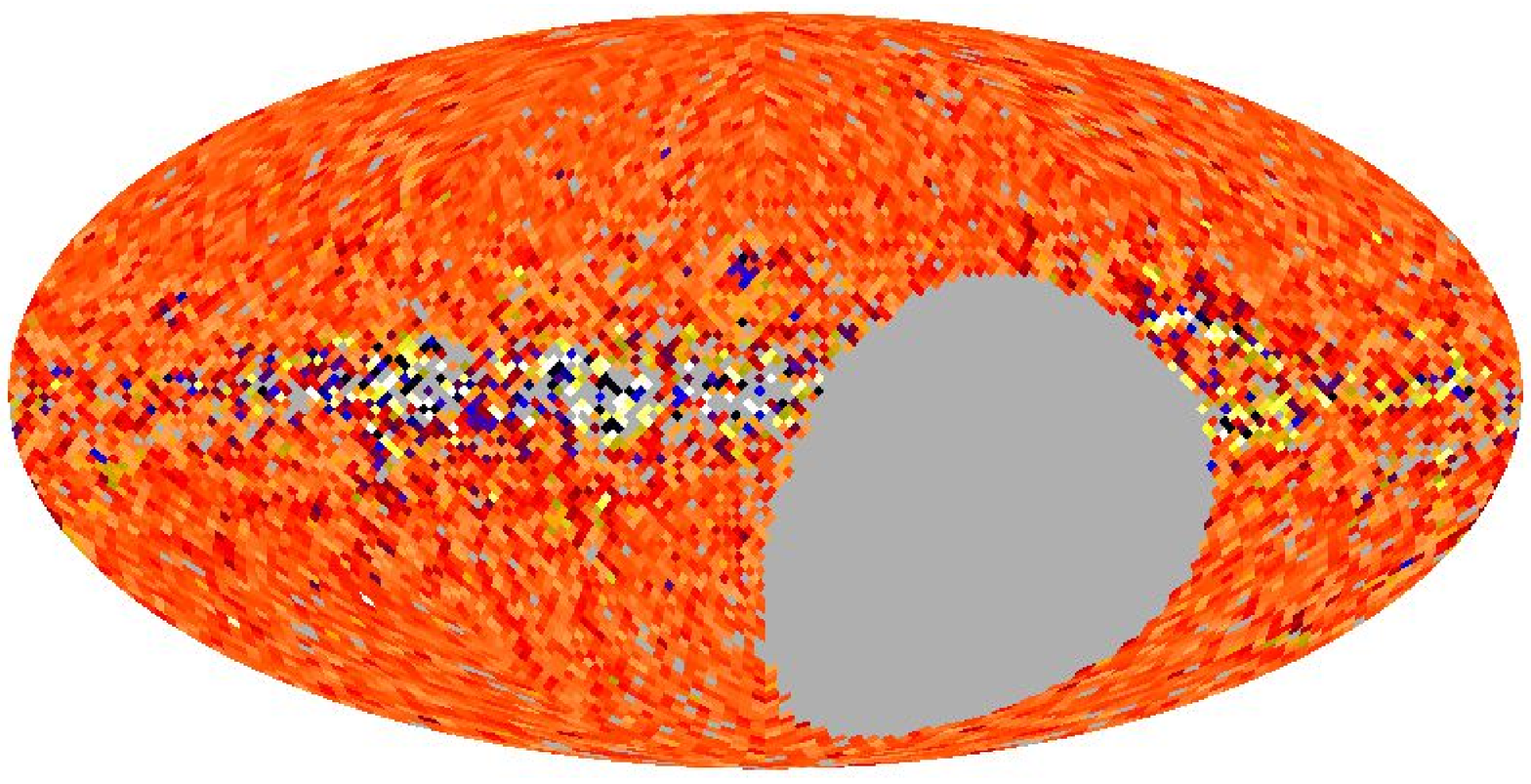}}\\
\subfigure[CS+GF+N($\sigma=1.0$ rad m$^{-2}$) with GF removal in $3^\circ$]
	{\includegraphics[width=0.98\textwidth]{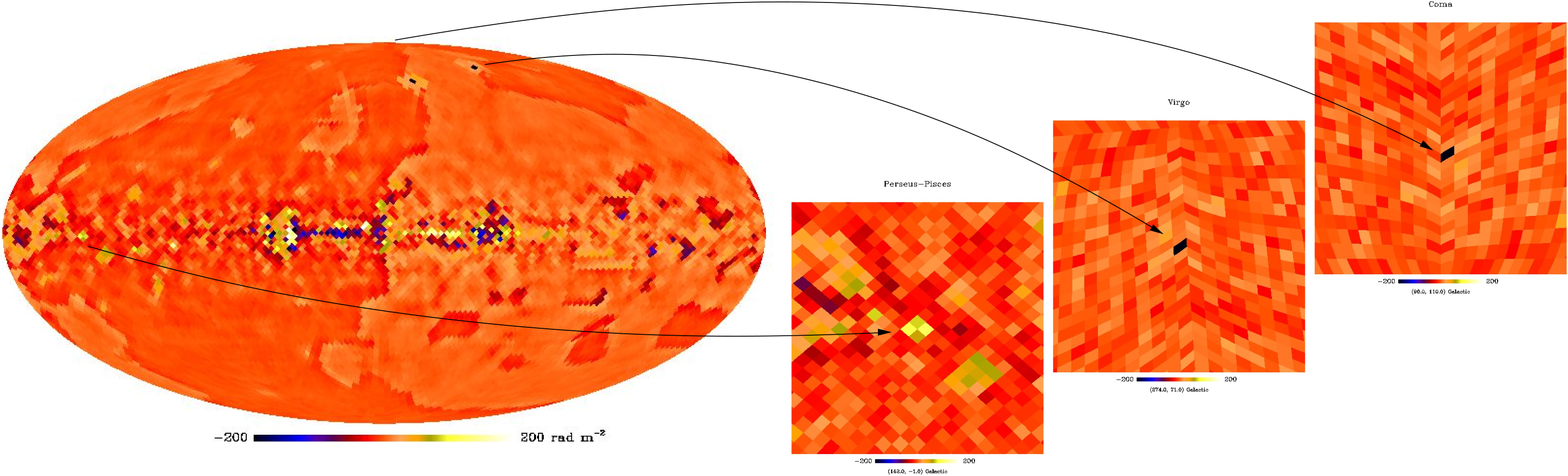}}

\caption{Full sky maps in galactic coordinates for the total synthetic
  signal (left column) and for the observed RMs (right column). The
  first row shows the Galactic foreground (GF) map generated with the
  {\small HAMMURABI} code of \citet{2009A&A...495..697W} including the
  cosmological signal (CS) from the {\it MHD Gal} simulation with a
  {\it magnetic depth} of $z=1$ and an imprinted observational error
  (N) of $\sigma=10$ rad m$^{-2}$ compared with the plain RM data
  given by \citet{2009ApJ...702.1230T}. The second row shows the same
  maps but smoothed by 8$^{\circ}$ \citep[as in Fig. 4
    of][]{2009ApJ...702.1230T} and the third row  shows the resulting
  residual maps when foreground removal is applied (within 
  3$^{\circ}$). The lower left plot shows the former synthetic map
  where the noise was reduced to $\sigma=1$ rad m$^{-2}$, as it is
  expected for future observations. In the lower right we show $40^\circ\times40^\circ$ 
  wide close-ups of three prominent clusters in the simulation 
  (from left to right: Perseus-Pisces, Virgo and Coma).
   \label{fig:simulgalaxy}}
\end{figure*}

\begin{figure*}
\includegraphics[width=0.46\textwidth]{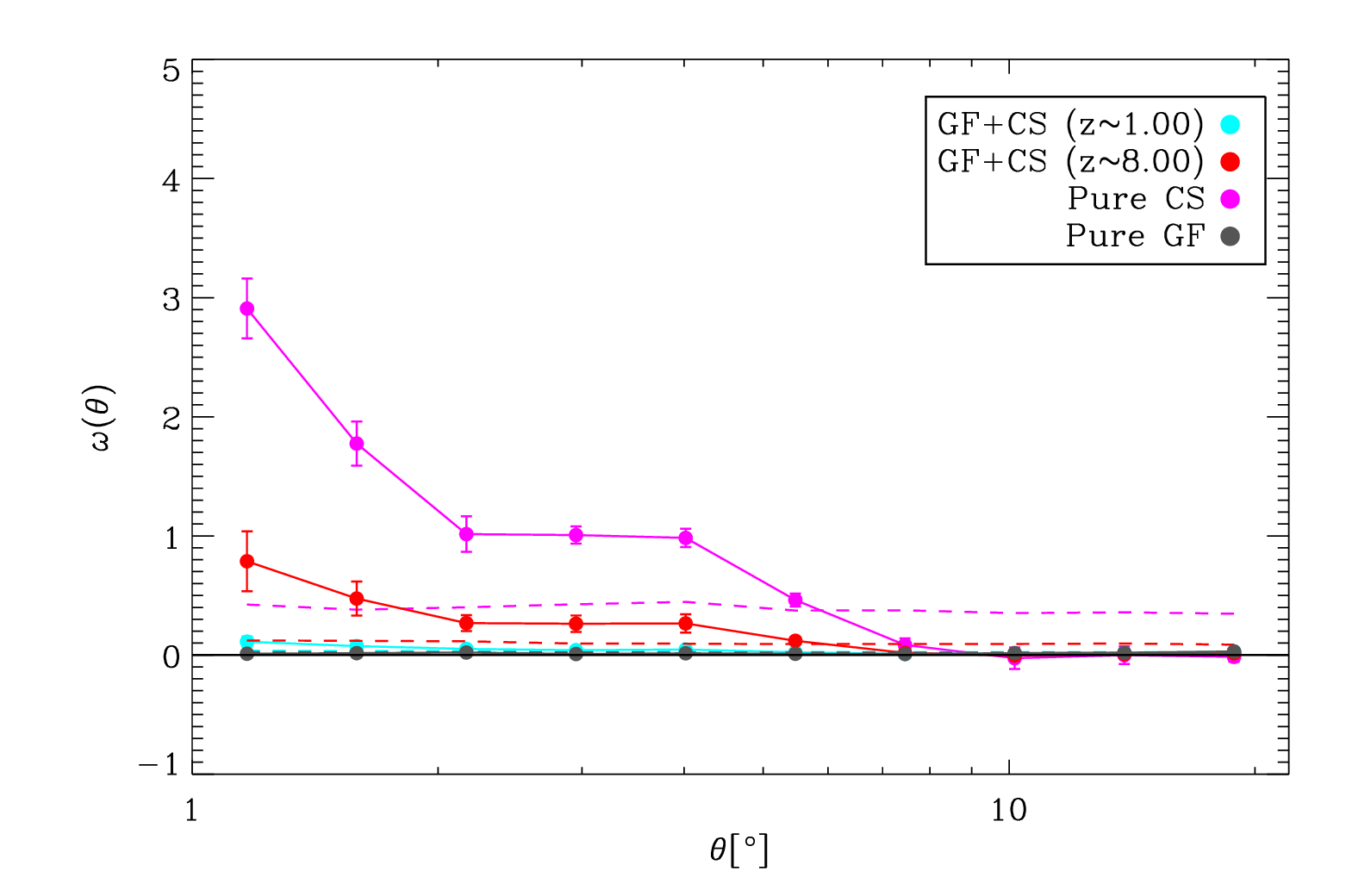}
\includegraphics[width=0.46\textwidth]{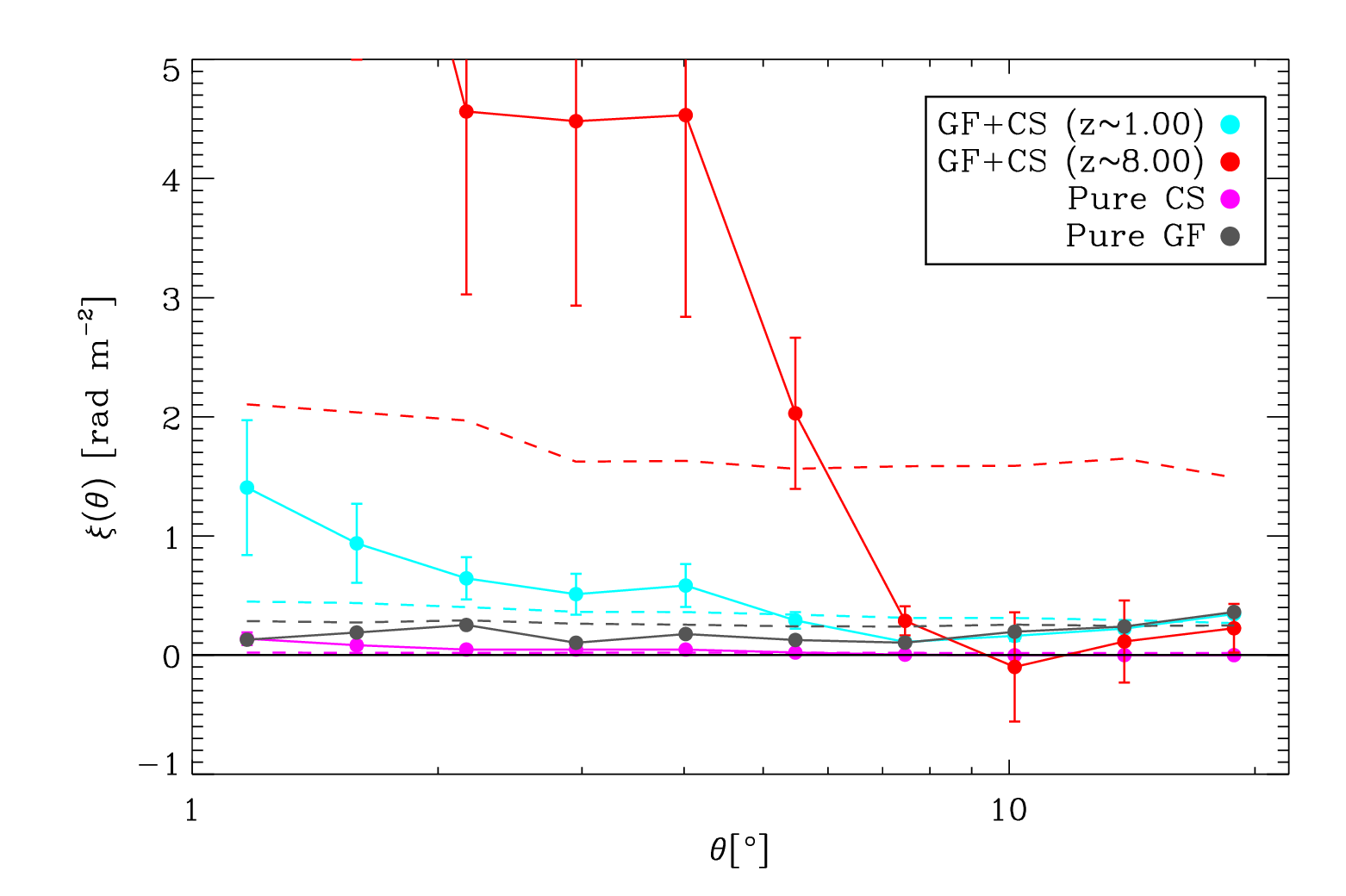}
\caption{Angular cross-correlation function for the combined cosmological signal (CS) from the {\it MHD Gal} 
simulation and the Galactic foreground (GF), including the foreground removal in $3^\circ$ as described in the text. Shown is the signal using both estimators as presented in the text (see Eqs. \ref{omega_est} and \ref{xi_est}).\label{fig:cross_agv4}}
\end{figure*}

In recent years, different models for the Galactic magnetic field were 
proposed \citep[e.g.][]{2006ApJ...642..868H, 2007ApJS..170..335P, 2008ICRC....2..223J, 2008A&A...477..573S}.
To estimate the influence of the GF on the cosmological cross-correlations 
we produced a synthetic map of the RM signal expected for our galaxy using the publicly available code 
{\small HAMMURABI} \citep{2009A&A...495..697W}, where we made use of the Galactic magnetic model given 
by \citet{2008A&A...477..573S}. 
The original model was constructed to give a good representation of the 
synchrotron emission of the Milky Way but, by missing possible reversals within the model magnetic 
field, it overproduces the RM signal by a significant factor. We therefore scaled the 
original model down to obtain a better representation of the observed RMs. We also 
note that such reversals could lead to significant small scale structures in the RM signal
due to the GF, as shown by \citet{2009A&A...507.1087S}. Such fluctuations could 
significantly compromise the cosmological signal, as they would be present on scales smaller than the one used to 
filter the GF. However, we do not currently include this effect in our foreground model. 

In Fig. \ref{fig:simulgalaxy} 
we show the obtained RM map models (left column) compared to the observed RM signal (right column)
taken from \citet{2009ApJ...702.1230T}. From top to 
bottom we show the original GF model with noise and the RM dataset, a smoothed version of the maps (within 8$^{\circ}$), and the residuals
when applying the foreground subtraction as described above for 3$^{\circ}$. All the synthetic maps are 
imprinted with an observational error of $\sigma=10$ rad m$^{-2}$. The last row shows the
synthetic residual map when reducing the noise level to $\sigma=1$ rad m$^{-2}$ as expected for future 
instruments. The close-ups show the remaining signal of prominent galaxy clusters in the residual maps.
Note that the signal of other prominent clusters, lying behind the Galactic plane, are not longer visible after 
the foreground subtraction was applied. 
As expected, when adding such a large, plain foreground signal to the cosmological one, the cross-correlation 
function vanishes. Therefore, we also applied the GF removal technique described before. The 
results can be seen in Fig. \ref{fig:cross_agv4}, 
where the angular cross-correlation function of the combined maps for both estimators is shown.
For comparison, we show the expected signal from the plain {\it MHD Gal} simulation (e.g. assuming a very
small {\it magnetic depth} of $z=0.03$), the GF signal alone, and the {\it MHD Gal} combined
with the GF signal for a large {\it magnetic depth} (i.e. $z=1.03$), as well as for a extreme
{\it magnetic depth} corresponding to $z\approx8$. In all cases we applied the foreground removal using 
a radius of $3^{\circ}$. Even for the extreme case of {\it magnetic depth}, despite the foreground removal
applied, the normalized estimator $\omega_{RM}$ drops further by a factor of $\sim3$ when adding the GF, 
and drops by a factor of $\sim30$ for the most optimistic cosmological signal. On the contrary, 
the unnormalized estimator $\xi_{RM}$ turns out to be quite insensitive to the foreground 
provided that the removal technique is applied. The combined signal still corresponds roughly to the original, 
cosmological one, as can be seen when comparing with Fig. \ref{fig:cross_depth}. 

We conclude, that although the normalized estimator $\omega_{RM}$ in principle contains
a much more unbiased and reliable imprint of the cosmological magnetization, once GF 
and observational noise are added, the underlying cosmological signal is completely lost.
In constrast, the unnormalized estimator $\xi_{RM}$ is relatively insensitive to the GF and to the noise. 
However, as seen before, the interpretation of its amplitude and shape is extremely challenging, as 
it is quite biased by the underlying {\it magnetic depth} probed by the redshift distribution of the 
radio sources used in the RM measurements.


\section{The simulated observational cross-correlations: an example}  \label{sec:res}

To study if it is possible to infer the underlying cosmological signal through an observational
process which includes GF (and its removal technique) as well as measurement errors by 
cross-correlating the $|{\rm RM}|$ signal with the galaxy density we assume:

\begin{itemize} 
\item An optimistic {\it magnetic depth} of the universe of $z=1.03$;
\item A GF according to the model presented in Section \ref{hammu} together with the foreground 
subtraction technique presented in Section \ref{for_rem} using the mean value of the $|{\rm RM}|$ map within 3$^{\circ}$;
\item A measurement error distribution consistent with a Gaussian having $\sigma_{\rm RM}=1$ rad m$^{-2}$, i.e. a dispersion 
similar to the magnitude of typical errors achievable by future instruments.
\end{itemize}

\subsection{Piling up the signal}
 
\begin{figure*}
\includegraphics[width=0.46\textwidth]{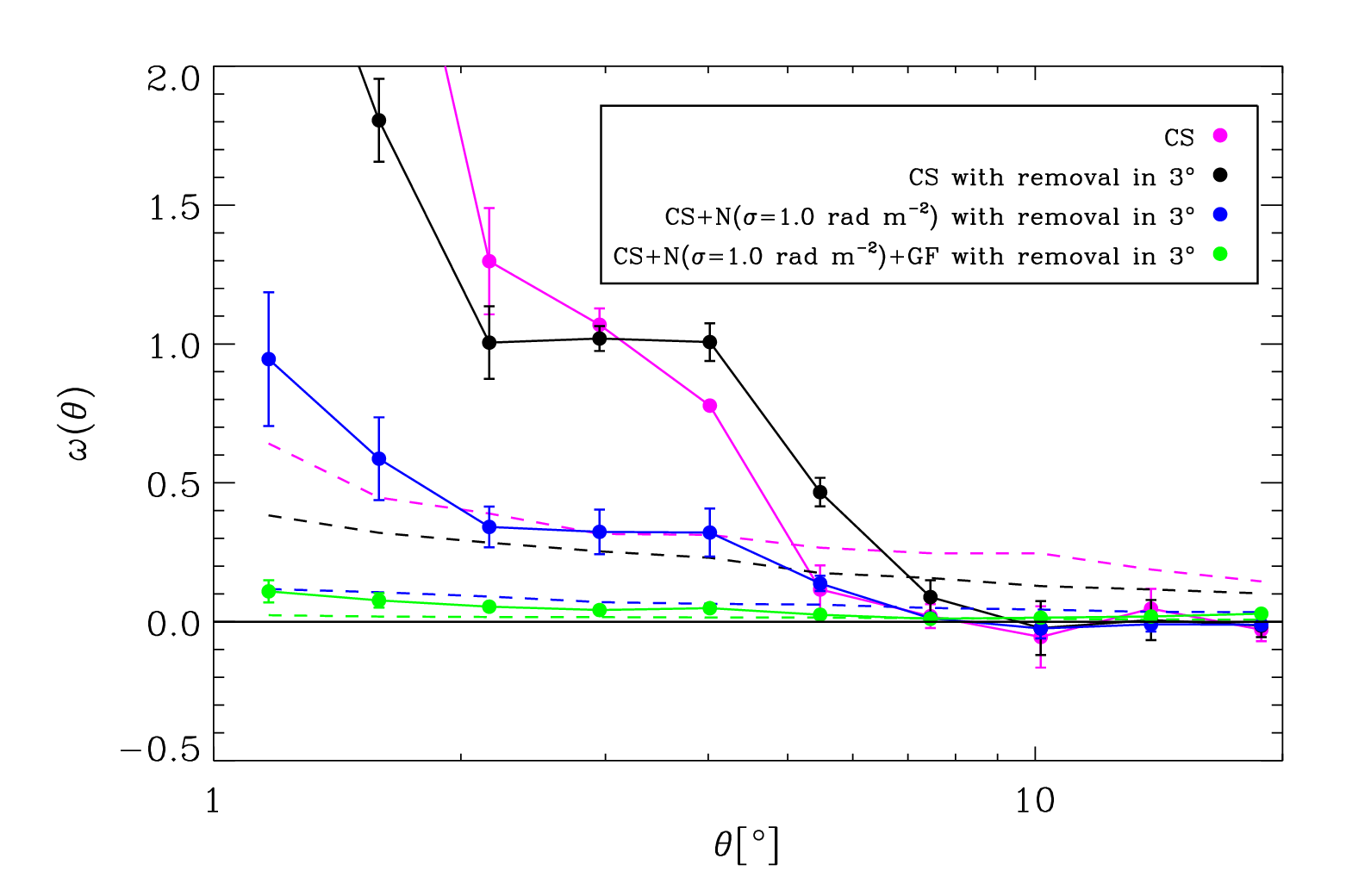}
\includegraphics[width=0.46\textwidth]{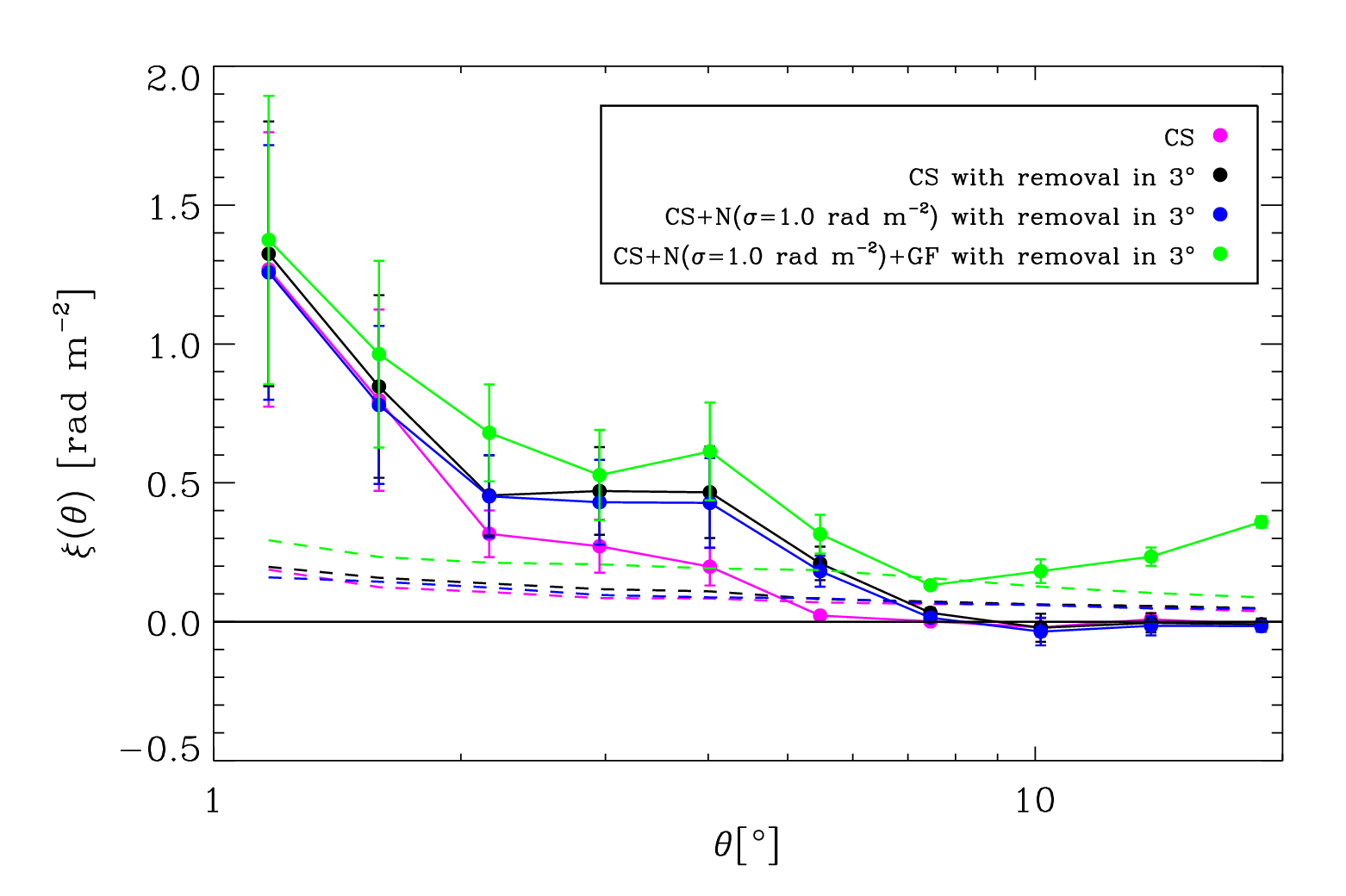}
\caption{Changes in the cross-correlation functions for the two estimators $\omega_{RM}$ (left panel) and
 $\xi_{RM}$ (right panel) when gradually including all described steps to the cosmological signal (CS) using the 
 {\it MHD Gal} simulation with a{\it magnetic depth} of the universe of $z=1.03$. GF and N stand for Galactic 
 foreground and observational noise respectively.\label{fig:cross_agv5}}
\end{figure*}

Fig. \ref{fig:cross_agv5} shows how the resulting signal changes when gradually adding all
effects described before, using the {\it MHD gal} model. As already seen in the individual steps
above, the normalized estimator $\omega_{RM}$ gives a much more significant signal, but its amplitude 
and shape suffers dramatically from the inclusion of noise and addition of the GF 
(despite of the substraction technique applied). On the other hand, the unnormalized estimator $\xi_{RM}$ 
gives a much less significant signal only mildly changed by all the contributions to 
the total signal, mainly at larger distances.

\subsection{Seeing different cosmic magnetizations}

\begin{figure*}
\includegraphics[width=0.46\textwidth]{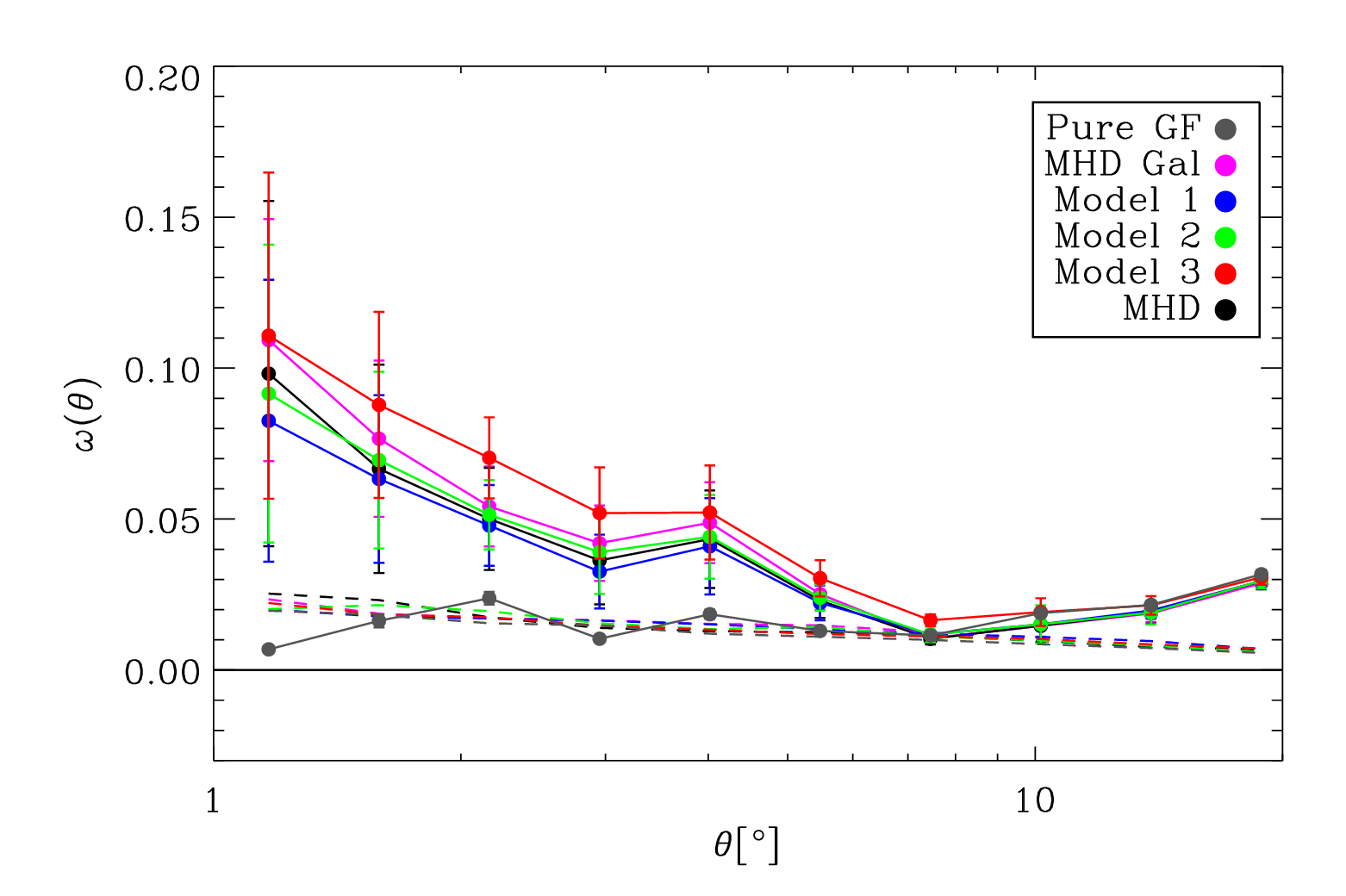}
\includegraphics[width=0.46\textwidth]{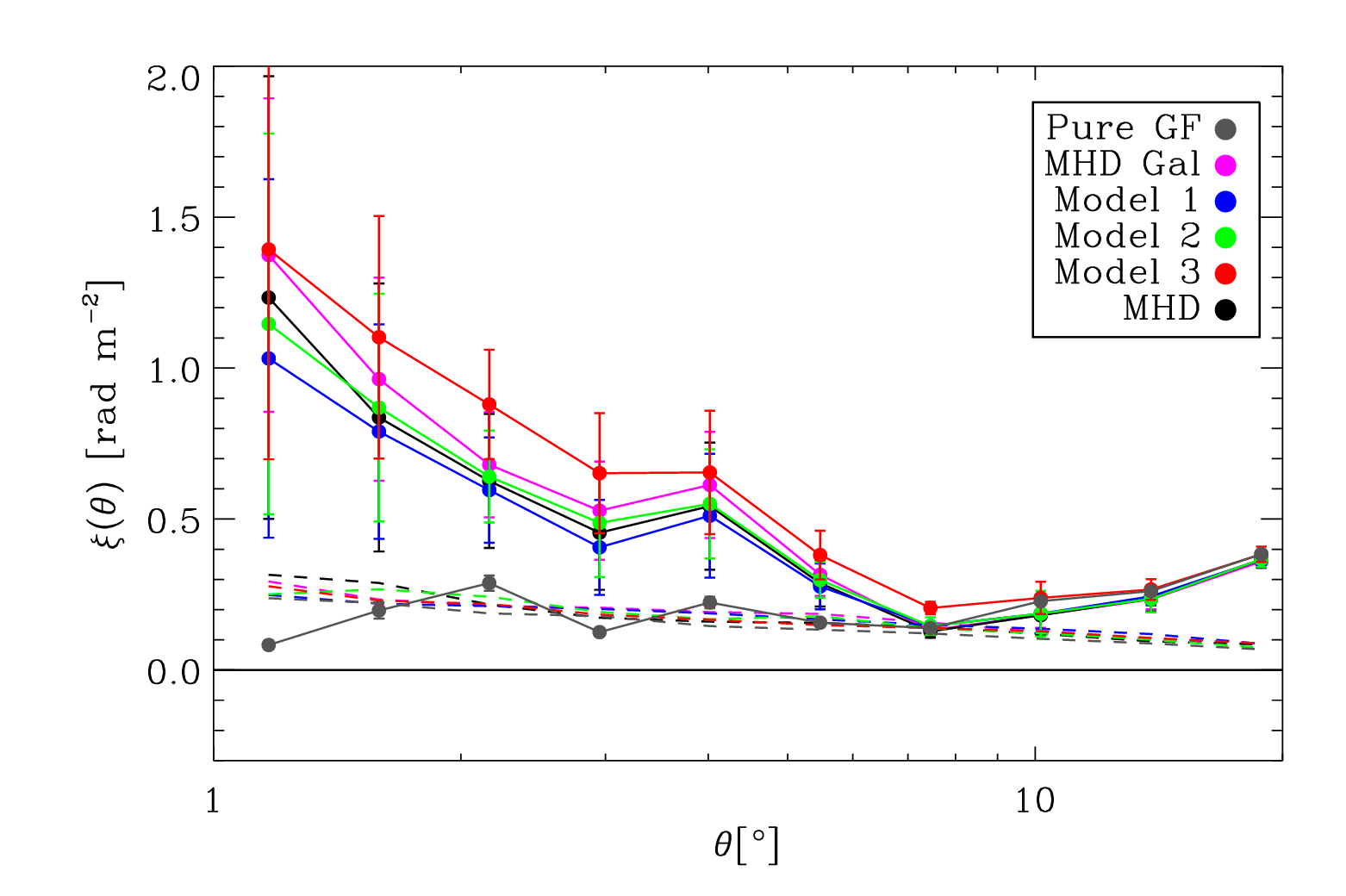}
\caption{Cross-correlation functions for the different magnetic field models, using an optimistic {\it magnetic depth} for 
the cosmological signal (CS) (corresponding to a universe magnetized out to $z=1.03$), taking into account the Galactic foreground (GF), 
assuming a Gaussian noise (N) with $\sigma_{\rm RM}=1$ rad m$^{-2}$ (as expected for the next generation of instruments)
and applying the foreground subtraction in $3^\circ$. Left panel shows the result for normalized estimator $\omega_{RM}$
whereas the right panel shows the result for the unnormalized estimator $\xi_{RM}$.\label{fig:cross_agv6}}
\end{figure*}

Fig. \ref{fig:cross_agv6} summarizes the results for such a combination of contributions to the total signal 
for our five different magnetic field models. The signal for the normalized 
estimator $\omega_{RM}$ (left panel) is reduced by a factor of $\sim50$ and the shape does not 
represent the underlying magnetic field models as well as when applied to the
cosmological signal itself (e.g. compare with Fig. \ref{fig:cross_models2}). 
The amplitude of the unnormalized estimator $\xi_{RM}$
corresponds to the underlying cosmological signal, but here the original ordering due to
the magnetic field models is no longer present. In general, for both estimators, the significance 
of the total signal is only marginal and the differences between the different magnetic field models 
lie far inside the error bars. Note that if we would only consider the {\it null} signal and ignore 
the errors from the magnetic field realization both estimators would give highly significant detections. 
The errors coming from the magnetic field realizations are not accessible
from the observations and, therefore, the observationally obtained significance of the signal can be 
misleading unless compared to detailed simulations. As well, both errors can be significantly 
reduced by using higher number of RMs. The results can also be improved 
by avoiding the Galactic region (e.g. cutting the region of the maps lying inside $\pm 10^{\circ}$). In that case, the 
unnormalized estimator will strongly reduce the power excess seen at distances larger than $3^\circ$. Such a 
cut would also remove the artificial but significant signal seen for separations larger than 10$^{\circ}$ in both estimators.


\section{Conclusions} \label{sec:conc}

Using cosmological MHD simulations of the magnetic field in galaxy
clusters and filaments we evaluated the possibility to infer the magnetic
field strength in filaments by measuring cross-correlation functions
between RMs and the galaxy density field. 

We find that the shape of the cross-correlation function using the normalized estimator
$\omega_{RM}$ (in absence of any noise or foreground signal) nicely reflects the
underlying distribution of magnetic field within the large scale structure.
However, a very large number of lines of sight probed by RM measurements (much more than 
the 3072 used in this investigation) are needed to overcome the statistical noise
induced by the particular magnetic field realization within the cosmic structures, in order to 
distinguish between the wide range of models we used here. In general, the RM signal 
is strongly dominated by the denser regions (e.g. those populated by galaxy clusters and groups) 
and not by the low density ones, like filaments. On this point, the magnetic 
field associated with filaments already changes by several orders of magnitudes within the 
different models used here.

Aditionally, the normalized estimator $\omega_{RM}$ is extremely sensitive to measurement errors
and to the presence of the GF (despite attempts to remove it by subtracting 
a smoothed map). It is fair to say that given the current measurement errors in the available RMs and our 
knowledge of the GF, present studies cannot determine the magnetization magnitude of the 
Universe based only on the cross-correlation $\omega_{RM}$, whatever the significance of the 
measured signal is.
On the contrary, the shape of the unnormalized estimator $\xi_{RM}$ (the same as used by Lee et al. 2009) is 
relatively insensitive against the presence of measurement errors for the RMs and for the presence of 
the GF (as long as the described removal technique is used). Its amplitude, however, 
is quite strongly affected by measurement uncertainties. Current measurement errors (as for example those 
inherited by the Taylor's published sample) suppress the signal by a significant amount in such a 
way that it is impossible to relate the amplitude of the cross-correlation function to the underlying magnetization of the 
the large scale structure. However, we expect that future radio telescopes will be able of reaching 
error magnitudes of order of 1 rad m$^{-2}$ that could make the correction of the signal possible. 

Unfortunately, 
this estimator does not nicely encode in its shape the details of the magnetization of the large scale 
structure and, especially, its amplitude is extremely sensitive to the {\it magnetic depth} of the Universe. 
Therefore, any interpretation of an observed signal is limited by our knowledge of the redshift distribution 
of the sources (towards the RM signals measured), as well as by our knowledge of the distribution and evolution of 
the cosmic universal magnetization. Future observational data will help to put better constraints
on theoretical models for the origin of cosmological magnetic fields which, in return, can be implemented
in next generation of MHD cosmological simulations in order to draw a self-consistent picture that can 
be compared against observations.

In summary, we conclude that current RM observations cannot constrain the amplitude and distribution
of magnetic fields within the large scale structure. On the other hand, future datasets, based on a larger number
of observations with more accurate RMs, might be able to shed light on the magnetic field
distribution and evolution within these structures. However, very detailed 
model predictions are needed in order to compare with any observed cross-correlation signal. It will be a quite demanding task for
future cosmological simulations to provide detailed enough information of the large scale structure 
magnetization process within a large enough volume to produce useful templates of such correlation 
functions which can then be compared directly to the observations.


\section*{acknowledgements}
The authors would like to thank Andr\'e Waelkens for making the {\small HAMMURABI} code publicly available 
and for generating the RM synthetic map of the Galactic foregound.
F.~S.~acknowledges partial support by PICT Max Planck 245 (2006) of the Ministry of Science and Technology (Argentina). 
S.~E.~N.~acknowledges the support of the DAAD (Deutscher Akademischer Austausch Dienst). 
K.~D.~acknowledges the support of the DFG Priority Programme 1177.

\bibliographystyle{mn2e}
\bibliography{master_RM}

\newpage

\appendix
\end{document}